\documentclass[12pt]{article}
\pdfoutput=1

\usepackage{amssymb, amsmath, amsfonts}
\usepackage{mathtools}

\usepackage{graphicx}

\usepackage{booktabs}
\usepackage{needspace}

\usepackage{setspace}
\numberwithin{equation}{section}
\usepackage{tocloft}

\def\be{\begin{equation}}
\def\ee{\end{equation}}
\renewcommand{\title}[1]{\vbox{\center\LARGE{#1}}\vspace{5mm}}
\newcommand{\address}[1]{\vbox{\center\em#1}}

\newcommand{\bra}[1]{\langle #1|}
\newcommand{\ket}[1]{|#1\rangle}
\newcommand{\braket}[2]{\langle #1|#2\rangle}
\newcommand{\Hcal}{\mathcal{H}}

\newcommand{\Jcal}{\mathcal{J}}
\newcommand{\BH}{\mathrm{BH}}
\newcommand{\AdS}{\mathrm{AdS}}
\newcommand{\avg}[1]{\overline{#1}}
\newcommand{\tfd}{\mathrm{TFD}}
\newcommand{\ghz}{\mathrm{GHZ}}
\newcommand{\pets}{\mathrm{PETS}}

\graphicspath{{./Fig_pdf/}}

\usepackage{xcolor}
\usepackage[pdftex,bookmarks=true,colorlinks,linkcolor=red,urlcolor=blue,citecolor=blue]{hyperref}
\usepackage{cite}

\begin{document}

\begin{titlepage}
\vspace{.5cm}

\begin{center}
\hfill \\
\hfill \\
\vskip 1cm

\title{\boldmath Constructing and Probing \\Three-Boundary Booklet Geometries}
\vskip 0.5cm
{Yan Liu}\footnote{Email: {\tt yanliu@buaa.edu.cn}},
{and Shaodong Zhou}\footnote{Email: {\tt zhoushaodong@buaa.edu.cn}}

\address{Department of Space Science, Beihang University, Beijing 100191, China}

\address{Peng Huanwu Collaborative Center for Research and Education, \\ Beihang University, Beijing 100191, China}

\end{center}

\vskip 1.5cm

\begin{abstract}
\noindent
Partially entangled thermal states (PETS) provide semiclassical families of black-hole microstates prepared through heavy shell insertions. We extend this shell-state
framework and propose three-boundary partially entangled thermal states ($\pets_3$) by replacing the two-index insertion data with a trivalent junction tensor. The corresponding bulk three-boundary booklet geometry is constrained semiclassically by the Israel junction condition, allowing us to identify candidate symmetric and mixed-orientation branches. We verify the self-consistency of this construction by counting states in the symmetric all-black-hole sector. Using a heavy-operator probe, we further estimate the detection-to-universal ratio to be $Z_D/Z_U \simeq 3$, twice the two-boundary value of $3/2$. These results suggest that a fine-tuned probe localized on one boundary can test the proposed three-boundary shell state and potentially distinguish it from its two-boundary counterpart.

\end{abstract}

\vfill

\end{titlepage}

\begingroup
\hypersetup{linkcolor=black}
\setlength{\cftbeforesecskip}{10pt}
\setlength{\cftbeforesubsecskip}{5pt}
\setcounter{tocdepth}{2}
\tableofcontents
\endgroup

\setcounter{footnote}{0}

\section{Introduction}
\label{sec:Introduction}

The Bekenstein-Hawking entropy, $S_{\rm BH}=A/(4G)$, suggests that a black hole with fixed macroscopic parameters has approximately
$e^{S_{\rm BH}}$ microstates, where $A$ is the horizon area \cite{Bekenstein1973,Hawking1975}. In AdS/CFT, a microscopic account of this entropy therefore requires sufficiently many independent CFT states with the same semiclassical black-hole exterior. The central difficulty is that distinct semiclassical interior geometries do not correspond to orthogonal quantum states. Consequently, simply counting such geometries may overestimate the number of independent black-hole microstates.

Partially entangled thermal states (PETS) provide a controlled setting in which to study this problem. They were first introduced as operator-inserted purifications in SYK \cite{Goel2019} and were subsequently realized as semiclassical black-hole microstates prepared by heavy dust-shell insertions in Euclidean gravity \cite{Balasubramanian2024,Balasubramanian2024b,Climent2024,GengJiang2025}. The shell has mass $m\sim {\cal O}(1/G_N)$ and its trajectory obeys the Israel junction conditions \cite{Israel1966}. Each PETS is a pure state, while its one-sided observables agree approximately with thermal expectation values only after coarse-graining and under appropriate eigenstate thermalization hypothesis (ETH) assumptions \cite{DAlessio2016}. Nonperturbative wormhole saddles provide access to the averaged overlap moments between distinct shell states. These overlaps determine the rank of the corresponding Gram matrix and, therefore, the number of linearly independent states in the semiclassical family \cite{Balasubramanian2024}.

A natural question arises: what is the multipartite generalization of this framework? It has been argued that purely GHZ-like entanglement is incompatible with an entropy inequality satisfied by a class of time-symmetric holographic states with smooth manifold bulk duals \cite{BalasubramanianKang2025}. A possible way to evade this obstruction is the booklet wormhole, a non-manifold geometry in which several bulk pages meet at a common interface \cite{JiangLiu2025}. The matter-junction structure and Lorentzian implications have also been explored \cite{JiangLiu2026}. The local gravitational matching of the pages is governed by multiway junction conditions \cite{Shen2024,Shen2024b}, which have also been realized explicitly in JT gravity \cite{Shen2026}. Related, though not identical, field-theoretic constructions involving junctions, interfaces, and network-like spaces have been explored in \cite{LiuWang2025,Banerjee2026,GuoMiao2026,GuoMiao2026b,Chakraborty2026}.
These developments motivate a three-boundary generalization of the shell-state construction.

In this paper, we propose a class of three-boundary partially entangled thermal states, denoted by $\pets_3$. We take a thermal GHZ state as the undeformed reference state and replace its diagonal junction tensor with general three-index junction amplitudes dressed by independent Euclidean filters. Motivated by conventional ETH and the statistics of multi-index OPE coefficients \cite{DAlessio2016,deBoer2024}, we introduce an ETH-inspired statistical ansatz for these junction amplitudes. We then use the three-page Israel junction conditions to identify candidate semiclassical geometries and
classify their admissible branches.
Compared with the two-boundary construction, the three-page booklet junction can permit configurations with unequal AdS radii (hence different central charges). We further examine the consistency of the construction by studying the Gram matrix and counting the resulting independent states. Finally, we extend the heavy-shell diagnostic of hidden preparation data \cite{BalasubramanianYildirim2025a}, using the associated nonperturbative path-integral framework \cite{BalasubramanianYildirim2025b}. Under the microcanonical factor-counting prescription described below, the all-black-hole $\pets_3$ candidate gives $Z_D/Z_U\simeq3$, compared with $3/2$ for the corresponding two-boundary state. This difference suggests an in-principle diagnostic of the number of entangled boundaries.

The paper is organized as follows. Sec.~\ref{sec:PETS3} introduces the thermal GHZ reference state, defines the junction-tensor construction of $\pets_3$, and discusses the conditions under which its one-boundary reduced states are thermal after coarse-graining. Sec.~\ref{sec:Israel} derives the three-page booklet Israel conditions and classifies the corresponding candidate saddle branches. Sec.~\ref{sec:counting} analyzes the rank of the Gram matrix for a generic family of three-index junction states. Sec.~\ref{sec:probes} applies the heavy-operator diagnostic and estimates the detection ratio in the all-black-hole sector. We conclude in Sec.~\ref{sec:conclusion}.

\section{Constructing $\pets_3$}
\label{sec:PETS3}

Heavy shell insertions provide semiclassically controlled families of AdS black-hole microstates \cite{Balasubramanian2024,Balasubramanian2024b,Climent2024,GengJiang2025}. In this section, we generalize this framework to three boundaries and propose a three-boundary partially entangled thermal state ($\pets_3$).

\subsection{Two views of PETS}
\label{sec:two-views}

We begin by reviewing two complementary descriptions of PETS. In the Euclidean path-integral description, a PETS is prepared by two Euclidean evolution segments separated by a heavy operator insertion. Semiclassically, the insertion creates a massive shell whose trajectory is determined by the gravitational junction conditions and can evolve into the black-hole interior. In the Hilbert-space description, the same construction can be viewed as a deformation of the thermofield-double (TFD) state. The canonical density matrix $\rho_\beta=e^{-\beta H}/Z(\beta)$ admits the TFD purification \cite{Maldacena2003}
\be
\ket{\tfd}_\beta
=
\frac{1}{\sqrt{Z(\beta)}}
\sum_n e^{-\beta E_n/2}\ket{n}_L\ket{n}_R .
\label{eq:TFD}
\ee
Tracing out either boundary gives $\rho_\beta$ on the remaining boundary. A PETS deforms this reference state by inserting a heavy operator $O_{\bf m}$, where ${\bf m}$ labels the shell data. In the energy basis, the state takes the form \cite{Balasubramanian2024}
\be
\ket{\Psi_{\bf m}^{(2)}}
=
\frac{1}{\sqrt{Z_{\bf m}^{(2)}}}
\sum_{i,j}
e^{-\tilde\beta_L E_i/2}
e^{-\tilde\beta_R E_j/2}
(O_{\bf m})_{ij}
\ket{E_i}_L\ket{E_j}_R .
\label{eq:PETS2}
\ee
The factors $e^{-\tilde\beta_L E_i/2}$ and $e^{-\tilde\beta_R E_j/2}$ represent Euclidean evolution along the two boundaries, while $(O_{\mathbf m})_{ij}$ encodes the shell-insertion data. When $O_{\mathbf m}$ is replaced by the identity and $\tilde\beta_L+\tilde\beta_R=\beta$, Eq.\ \eqref{eq:PETS2} reduces to the TFD state in \eqref{eq:TFD}.

The generalized ETH ansatz for the shell-operator matrix elements takes the form \cite{Balasubramanian2024,DAlessio2016}
\be
(O_{\mathbf m})_{ij}
=
e^{-S(\bar E)/2}\,
g_{\mathbf m}(\bar E,\omega)^{1/2}\,
R_{ij}^{({\mathbf m})},
\qquad
\bar E=\frac{E_i+E_j}{2},\quad
\omega=\frac{E_i-E_j}{2}.
\label{eq:ETH}
\ee
Here $R_{ij}^{(\mathbf m)}$ is a pseudorandom matrix whose fine-grained phases average to zero in coarse-grained observables, while the smooth envelope $g_{\mathbf m}$ is fixed by the semiclassical shell geometry. Under the ETH and coarse-graining assumptions, one-sided simple observables approximately agree with their thermal expectation values, even though the full PETS remains
pure and retains information about its preparation.

These two descriptions suggest a route to the three-boundary construction:
the TFD reference state is replaced by a thermal GHZ reference state, and the two-index insertion data are replaced by three-index amplitudes associated with a trivalent junction.

\subsection{Three-boundary junction states}
\label{sec:GHZ}

We now generalize the descriptions for PETS above. As a reference state for the three-boundary construction, we consider a thermal GHZ state whose reduction to any single boundary is thermal. We then deform this diagonal reference state by replacing its junction tensor with general three-index amplitudes.

Starting from $\rho_\beta=\sum_n p_n\ket{n}\bra{n}$ with $p_n=e^{-\beta E_n}/Z(\beta)$, we define
\be
\ket{\ghz}_\beta
=
\sum_n\sqrt{p_n}\,
\ket{n}_1\ket{n}_2\ket{n}_3
=
\frac{1}{\sqrt{Z(\beta)}}
\sum_n e^{-\beta E_n/2}
\ket{n}_1\ket{n}_2\ket{n}_3 .
\label{eq:GHZ}
\ee
This definition assumes three isomorphic Hilbert spaces with aligned energy eigenbases. Tracing out any two factors gives $\rho_\beta$ on the remaining boundary. Unlike a generic state in $\Hcal_1\otimes\Hcal_2\otimes\Hcal_3$, however, the thermal GHZ state is supported only on the diagonal subspace with aligned energy labels.

The thermal GHZ state is relevant here because it motivates a bulk geometry that lies outside the usual class of smooth single-manifold constructions. A class of time-symmetric holographic states with smooth manifold bulk duals obeys $\tfrac12 R^{(3)}\ge GM^{(3)}$, where $R^{(3)}=S_R(A:B)-I(A:B)$ is the residual information and $GM^{(3)}$ is the genuine multi-entropy. By contrast, a GHZ state has $R^{(3)}=0<GM^{(3)}$ \cite{BalasubramanianKang2025}. A booklet geometry has been proposed as a possible bulk description of such a state \cite{JiangLiu2025}; because it is not a single smooth manifold, it lies outside the assumptions of the corresponding minimal-surface argument.

We now deform the thermal GHZ reference state \eqref{eq:GHZ} by introducing a heavy trivalent insertion at the common junction. The insertion generally prepares a three-boundary junction state
\be
\ket{\Jcal_A}
\in
\Hcal_1\otimes\Hcal_2\otimes\Hcal_3 .
\label{eq:junction-state}
\ee
The label $A$ specifies the insertion and junction data, including the shell quantum numbers. We denote the three outgoing energy-basis amplitudes of the insertion by
\be
O_{ijk}^{(A)}
\equiv
\braket{E_i,E_j,E_k}{\Jcal_A} ,
\label{eq:O3}
\ee
where $\ket{E_i,E_j,E_k}\equiv\ket{E_i}_1\ket{E_j}_2\ket{E_k}_3$. Dressing the three legs of the insertion with independent Euclidean evolution factors gives the three-boundary state, namely $\pets_3$,
\be
\ket{\Psi_A^{(3)}}
=
\frac{1}{\sqrt{Z_A^{(3)}}}
\sum_{i,j,k}
e^{-\tilde\beta_1E_i/2}
e^{-\tilde\beta_2E_j/2}
e^{-\tilde\beta_3E_k/2}
O_{ijk}^{(A)}
\ket{E_i}_1\ket{E_j}_2\ket{E_k}_3 ,
\label{eq:PETS3state}
\ee
with normalization
\be
Z_A^{(3)}
=
\sum_{i,j,k}
e^{-\tilde\beta_1E_i-\tilde\beta_2E_j-\tilde\beta_3E_k}
|O_{ijk}^{(A)}|^2 .
\label{eq:Z3}
\ee
Throughout this construction, $O_{ijk}^{(A)}$ denotes a set of three-leg junction amplitudes, rather than the matrix elements of an operator acting within a single boundary Hilbert space.

The undeformed reference insertion tensor is
\be
O_{ijk}^{(0)}=\delta_{ij}\delta_{jk}.
\label{eq:ref-insertion}
\ee
The three Euclidean filters then combine into $\exp[-\tfrac12(\tilde\beta_1+\tilde\beta_2+\tilde\beta_3)E_i]$. When $\tilde\beta_1+\tilde\beta_2+\tilde\beta_3=\beta$, the normalization becomes $Z_0^{(3)}=Z(\beta)$, and the $\pets_3$ state \eqref{eq:PETS3state} reduces to the thermal GHZ state \eqref{eq:GHZ} \cite{JiangLiu2025}. For a generic value of $A$, however, $O_{ijk}^{(A)}$ need not remain diagonal in the three energy labels. Thus, the thermal GHZ state serves as the undeformed reference state; a generic $\pets_3$ state is not assumed to remain in the GHZ-diagonal sector.

To specify a tractable ensemble of deformed junction tensors, we introduce an ETH-inspired statistical ansatz. This is not the conventional ETH ansatz for matrix elements of a single-boundary operator. Rather, it is a modeling assumption for the three-index amplitudes, motivated by ordinary operator ETH, the shell-state construction, and related multi-index OPE statistics \cite{Balasubramanian2024,DAlessio2016,deBoer2024}. We define the smooth coarse-grained variance by
\[
W_A(E_i,E_j,E_k)
\equiv
e^{-S(\bar E)}g_A(E_i,E_j,E_k),
\qquad
\bar E\equiv\frac{E_i+E_j+E_k}{3},
\]
and parameterize the amplitudes as
\be
O_{ijk}^{(A)}
=
W_A(E_i,E_j,E_k)^{1/2}R_{ijk}^{(A)}
=
e^{-S(\bar E)/2}\,
g_A(E_i,E_j,E_k)^{1/2}\,
R_{ijk}^{(A)}.
\label{eq:ETH3}
\ee
Here $R_{ijk}^{(A)}$ is a pseudorandom three-index tensor, while the smooth spectral weight $W_A$ encodes the insertion and junction data.
We assume the coarse-grained two-point contraction
\be
\avg{R_{ijk}^{(A)}R_{i'j'k'}^{(B)*}}
=
\delta_{AB}\,\delta_{ii'}\,\delta_{jj'}\,\delta_{kk'}.
\label{eq:ETH3-correlator}
\ee
The overline denotes a coarse-grained or ensemble average over fine-grained phases that are not resolved
by the semiclassical gravitational description \cite{Balasubramanian2024}. Equation \eqref{eq:ETH3-correlator}, including the statistical independence
of different labels $A$, is an assumption defining the junction-tensor ensemble.

We next derive a sufficient condition for coarse-grained one-boundary thermality.
The $\pets_3$ state is pure in $\Hcal_1\otimes\Hcal_2\otimes\Hcal_3$,
but its reduced state on boundary 1 is
\be
(\rho_1)_{ii'}
=
\frac{
e^{-\tilde\beta_1(E_i+E_{i'})/2}
}{
Z_A^{(3)}
}
\sum_{j,k}
e^{-\tilde\beta_2E_j-\tilde\beta_3E_k}
O_{ijk}^{(A)}O_{i'jk}^{(A)*}.
\label{eq:rho1}
\ee
Assuming that the normalization factor is self-averaging at large entropy, $Z_A^{(3)}\simeq\avg{Z_A^{(3)}}$,
using \eqref{eq:ETH3} and \eqref{eq:ETH3-correlator}, we obtain
\be
\avg{(\rho_1)_{ii'}}
\simeq
\delta_{ii'}
\frac{
e^{-\tilde\beta_1E_i}\Xi_1^{(A)}(E_i)
}{
\avg{Z_A^{(3)}}
},
\label{eq:rho1-avg}
\ee
where
\be
\Xi_1^{(A)}(E_1)
=
\sum_{j,k}
e^{-\tilde\beta_2E_j-\tilde\beta_3E_k}
W_A(E_1,E_j,E_k)
\label{eq:Xi1}
\ee
is the effective spectral weight induced by tracing out boundaries 2 and 3. Equation~\eqref{eq:rho1-avg} is diagonal in the energy basis, but it is not automatically of Gibbs form. At large $N$, the sum defining $\Xi_1^{(A)}$ is expected to be dominated by a narrow saddle-point region. If $\log\Xi_1^{(A)}(E_1)$ can be linearized throughout the corresponding energy window,
\[
\log\Xi_1^{(A)}(E_1)\simeq c_1-\gamma_1E_1,
\]
then the averaged reduced state becomes
\be
\avg{\rho_1}
\simeq
\frac{e^{-\beta_{1,\rm eff}H_1}}{Z_1(\beta_{1,\rm eff})}
\label{eq:rho1-thermal}
\ee
with $\beta_{1,\rm eff}=\tilde\beta_1+\gamma_1$. Analogous expressions hold for boundaries 2 and 3, rendering each individual boundary thermally indistinguishable after coarse-graining. If the junction tensor and the Euclidean filters are permutation symmetric, the three effective temperatures coincide; in the general asymmetric construction, they need not. The argument above establishes an ensemble-averaged statement. We additionally assume concentration around this average in the large-entropy regime when interpreting a typical junction state as thermally indistinguishable by one-boundary coarse-grained observables.

\begin{figure}[h!]
\centering
\includegraphics[width=0.55\textwidth,trim=152pt 179pt 186pt 151pt,clip]{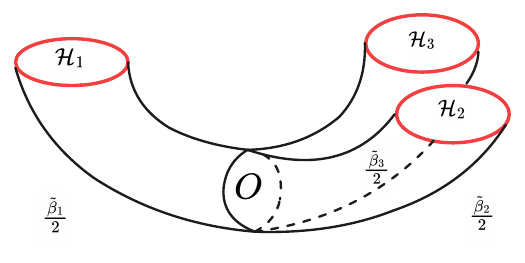}
\caption{\small Schematic Euclidean path integral preparing a $\pets_3$ state. The
three Euclidean legs meet at a common heavy junction insertion. The transverse
section of each leg represents the spatial sphere $S^{d-1}$, while its axial
direction represents Euclidean evolution.}
\label{fig:prepared-state}
\end{figure}

Having defined the state in the boundary Hilbert space, we now formulate a booklet-inspired Euclidean path-integral ansatz for its preparation, as shown in Fig.~\ref{fig:prepared-state}. Three Euclidean evolution segments of lengths $\tilde\beta_i/2$ terminate at a common heavy trivalent insertion. Each asymptotic boundary is associated with one CFT factor. Following the shell-state construction, the insertion is modeled as a spherically symmetric composite of $N_{\rm op}\sim{\cal O}(N^2)$ single-trace operators of dimension $\Delta\sim{\cal O}(1)$ distributed over an $S^{d-1}$. Its bulk backreaction produces a dust shell with inertial mass $m\sim{\cal O}(1/G_N)$, described semiclassically as a localized pressureless fluid with surface density $\sigma$. The multiway Israel junction conditions then provide the local gravitational matching conditions among the three bulk pages.

Notably, the spacetime here is no longer a geometry of the form $\mathbb{R} \times X$; it is not even a manifold in the conventional sense. We therefore treat each page as an ordinary Euclidean manifold and impose the multiway matching conditions along the shared shell worldvolume. This prescription defines a local semiclassical saddle ansatz, but it does not by itself determine the global sewing of the pages, the complete gravitational path-integral measure, or a Lorentzian continuation. In the next section, we derive these local matching conditions and analyze the resulting candidate branches.

\section{Three-Boundary Booklet Israel Conditions}
\label{sec:Israel}

We now examine whether the boundary states introduced in Sec.~\ref{sec:PETS3} admit semiclassical booklet descriptions. Each page is taken to be a Euclidean AdS--Schwarzschild or thermal-AdS region, and the three pages meet along a common codimension-one shell worldvolume $\mathcal W$. For standard canonical AdS boundary conditions, thermal AdS dominates below the Hawking--Page temperature, whereas the large Euclidean black hole dominates above it \cite{HawkingPage1983}.
Once the pages are coupled through a common shell, the configuration is governed by the Israel junction conditions.

The Euclidean action is taken to be
\be
I_E=
-\sum_{i=1}^{3}\frac{1}{16\pi G}
\int_{\mathcal M_i}\!\sqrt g\,(R-2\Lambda_i)
-\sum_{i=1}^{3}\frac{1}{8\pi G}
\int_W\!\sqrt h\,K_i
+I_{\rm shell}[h]+I_{\rm bdy,ct}.
\label{eq:action}
\ee
Here $I_{\rm bdy,ct}$ collects the asymptotic boundary, joint, and counterterm contributions, which do not modify the local matching equation. For simplicity, we take the Newton constants on the three pages to be equal. The multiway Israel junction conditions are
\be
h_{ab}^{(1)}=h_{ab}^{(2)}=\cdots=h_{ab}^{(n)}\equiv h_{ab},
\qquad
\sum_{i=1}^{n}\left(K_{ab}^{(i)}-K^{(i)}h_{ab}\right)
=-8\pi G S_{ab},
\label{eq:tensor-junction}
\ee
where $h_{ab}$ is the metric induced on $\mathcal W$, $K_{ab}^{(i)}$ is the extrinsic curvature computed with the chosen outward normal on page $i$, and $S_{ab}$ is the shell stress tensor.

To specialize these equations to a spherically symmetric shell, consider $n$ pages with metrics
\be
ds_i^2
=
f_i(r_i)d\tau_i^2
+\frac{dr_i^2}{f_i(r_i)}
+r_i^2d\Omega_{d-1}^2 ,
\qquad
f_i(r)=1+\frac{r^2}{L_i^2}-\frac{\mu_i}{r^{d-2}},
\label{eq:metric}
\ee
with $\mu_i=16\pi G_{d+1}M_i/[(d-1)\Omega_{d-1}]$. Thermal AdS corresponds to $M_i=0$, while $M_i>0$ describes an AdS--Schwarzschild page. Unless stated otherwise, we take $L_1=L_2=L_3\equiv L$. If the $L_i$ differ, the pages generally correspond to CFTs with different central charges, and the interpretation of a common shell area requires additional care.

Parametrize the shell trajectory on page $i$ by Euclidean proper time $T$: $r_i=r_i(T)$ and $\tau_i=\tau_i(T)$. Continuity of the induced metric gives
\be
f_i(r)\dot\tau_i^2+\frac{\dot r_i^2}{f_i(r)}=1,
\label{eq:ind-met}
\ee
where a dot denotes $d/dT$, together with
\be
r_1=r_2=\cdots=r_n\equiv r,
\label{eq:common-r}
\ee
because the shell has a common areal radius on all pages.
For a pressureless shell of conserved inertial mass $m$, the surface density scales as $\sigma=m/(\Omega_{d-1}r^{d-1})$. The angular component of Eq.~\eqref{eq:tensor-junction} then reduces to
\be
\sum_{i=1}^n f_i(r)\dot\tau_i
=\sum_{i=1}^{n} \epsilon_i\sqrt{f_i(r)-\dot r^2}
=
\frac{8\pi G_{d+1}m}{(d-1)\Omega_{d-1}r^{d-2}}
\equiv \kappa(r),
\label{eq:israel-n}
\ee
where $\epsilon_i=\pm1$ records the orientation
of the shell trajectory on page $i$.
For three pages $n=3$, Eq.\ \eqref{eq:israel-n} becomes
\be
\epsilon_1 \sqrt{f_1(r)-\dot r^2}
+\epsilon_2 \sqrt{f_2(r)-\dot r^2}
+\epsilon_3 \sqrt{f_3(r)-\dot r^2}
=
\kappa(r).
\label{eq:israel-3}
\ee
The square roots must remain real along the entire trajectory. We next analyze several classes of solutions to this local equation.

\subsection{Symmetric three-boundary geometries}
\label{sec:symmetric}

We first consider the symmetric case with $M_1=M_2=M_3\equiv M$, $L_1=L_2=L_3\equiv L$, and $\epsilon_1=\epsilon_2=\epsilon_3=+1$. Thus $f_1=f_2=f_3\equiv f$.
Equation \eqref{eq:israel-3} then reduces to $3\sqrt{f(r)-\dot r^2}=\kappa(r)$. Writing the radial equation as $\dot r^2+V_{\rm eff}(r)=0$ gives
\be
V_{\rm eff}(r)=\frac{\kappa(r)^2}{9}-f(r).
\label{eq:Veff-sym}
\ee
This branch closely parallels the radial shell dynamics of the two-boundary PETS construction \cite{Balasubramanian2024}.

We now analyze the behavior of $V_{\rm eff}$ for $d\ge3$. When $r\to0$, the $\kappa(r)^2$ term scales as $r^{-2d+4}$ and dominates the black-hole term in $f(r)$, giving
\be
V_{\rm eff}(r) \approx \frac{C}{r^{2d-4}} > 0,\qquad r \to 0,
\label{eq:Veff-zero}
\ee
with $C>0$. The positive divergence prevents the shell from reaching the origin. At large $r$,
\be
V_{\rm eff}(r) \approx - \frac{r^2}{L^2} \to -\infty, \qquad r \to \infty.
\label{eq:Veff-inf}
\ee
Consequently, a trajectory arriving from the asymptotic region can reach a minimum radius $r_*$, determined by $V_{\rm eff}(r_*)=0$, and return to the boundary. The Euclidean time elapsed on page $i$ is
\[
\Delta\tau_i
=2\int_{r_*}^{\infty}dr\,
\frac{\sqrt{f_i(r)+V_{\rm eff}(r)}}{f_i(r)\sqrt{-V_{\rm eff}(r)}}.
\]
The Euclidean preparation interval $\tilde\beta_i$ is related to the complete boundary period by
\be
\beta_i=\tilde\beta_i+\Delta\tau_i .
\label{eq:beta-rel}
\ee
For a black-hole page, $\beta_i$ must also agree with the period required for regularity at the Euclidean horizon. In the fully symmetric branch, $\Delta\tau_i$ and the regularity periods are identical on all three pages; therefore, this branch realizes the permutation-symmetric subfamily of the boundary construction in Sec.~\ref{sec:PETS3}. More general choices of the three Euclidean filters require an asymmetric junction solution.

\subsection{Mixed time orientations and unequal AdS radii}
\label{sec:mixed-L}

The three-page junction also permits mixed orientation signs. Such signs can allow cancellations among the square roots and thereby accommodate pages with different asymptotic data of unequal $L_i$.
For convenience, define $Y_i(r)\equiv f_i(r)+V_{\rm eff}(r)$.

Eq.\ \eqref{eq:israel-3} becomes
\be
\epsilon_1\sqrt{Y_1}
+\epsilon_2\sqrt{Y_2}
+\epsilon_3\sqrt{Y_3}
=\kappa.
\label{eq:israel-Y}
\ee
Repeated squaring eliminates the orientation signs and yields the quartic equation
\be
\left[
\left(\kappa^2-p\right)^2-4q\right]^2-64\kappa^2s =0,
\label{eq:quartic}
\ee
where $p=Y_1+Y_2+Y_3$, $q=Y_1Y_2+Y_1Y_3+Y_2Y_3$, and $s=Y_1Y_2Y_3$. Because the squaring procedure loses the orientation information and can introduce spurious solutions, every algebraic root must be substituted back into Eq.~\eqref{eq:israel-Y}. A physical branch must satisfy the original unsquared equation, $Y_i\geq0$, and $-V_{\rm eff}=\dot r^2\geq0$ along the entire trajectory.

The large-radius behavior imposes additional restrictions. For $d>3$,
\be
f_i(r) \simeq \frac{r^2}{L_i^2} + 1 - \frac{\mu_i}{r^{d-2}} + \cdots, \qquad \kappa(r) \simeq {\cal O}\!\left(\frac{1}{r^{d-2}}\right).
\label{eq:large-r-exp}
\ee
Suppose $V_{\rm eff}(r) = \alpha r^2 + \beta + o(1)$ as $r\to\infty$. Then
\be
Y_i = f_i + V_{\rm eff} \simeq (a_i + \alpha)\, r^{2} + (1+\beta) + \cdots, \qquad a_i \equiv \frac{1}{L_i^{2}}.
\label{eq:Yi-exp}
\ee
When $a_i + \alpha > 0$, the corresponding square root behaves as
\be
\sqrt{Y_i} \simeq \sqrt{a_i + \alpha} r + \frac{1+\beta}{2\sqrt{a_i+\alpha}}\frac{1}{r} + \cdots.
\label{eq:sqrt-exp}
\ee
The leading ${\cal O}(r)$ term in the junction equation must therefore obey
\be
\Big(\sum_i \epsilon_i\sqrt{a_i+\alpha}\Big) r = 0.
\label{eq:Or}
\ee
If all $\epsilon_i$ share the same sign, each term in this sum must vanish separately, i.e. $\sqrt{a_i+\alpha}=0$. This requires $a_1=a_2=a_3=-\alpha$ and hence $L_1=L_2=L_3$. This is precisely the symmetric case solved above.

By contrast, mixed orientations can satisfy Eq.~\eqref{eq:Or} through cancellations. For example, the leading condition associated with the sign pattern $(+,-,-)$ is
\be
\sqrt{a_1 + \alpha} - \sqrt{a_2 + \alpha} - \sqrt{a_3 + \alpha} = 0.
\label{eq:mixed}
\ee
The overall sign-reversed pattern obeys the same leading cancellation, although the original unsquared equation selects the admissible orientation at subleading order.

Squaring Eq.~\eqref{eq:mixed} twice gives
\be
\alpha_{\pm} = \frac{1}{3}\left(-\sum_i a_i \pm 2\Delta\right), \qquad
\Delta \equiv \sqrt{\sum_i a_i^{2} - \sum_{i<j} a_i a_j}.
\label{eq:alpha-sol}
\ee
The algebraic roots must still satisfy $a_i+\alpha\geq0$ and $\alpha\leq0$, the latter being required by $\dot r^2=-V_{\rm eff}\geq0$ asymptotically. The $\alpha_-$ root typically violates at least one of these conditions, while $\alpha_+$ can define a physical asymptotic branch. On the boundary $\alpha_+=0$, the leading condition becomes
\be
\frac{1}{L_1} = \frac{1}{L_2} + \frac{1}{L_3}.
\label{eq:L-relation}
\ee
For $d>3$ and for a nondegenerate coefficient of the $1/r$ term in Eq.~\eqref{eq:sqrt-exp}, the next-order cancellation gives $V_{\rm eff}\rightarrow-1$. When $d=3$, the right-hand side $\kappa={\cal O}(1/r)$ contributes at the same order, and the subleading asymptotic analysis must be modified.

\begin{figure}[h!]
\centering
\includegraphics[width=0.5\textwidth]{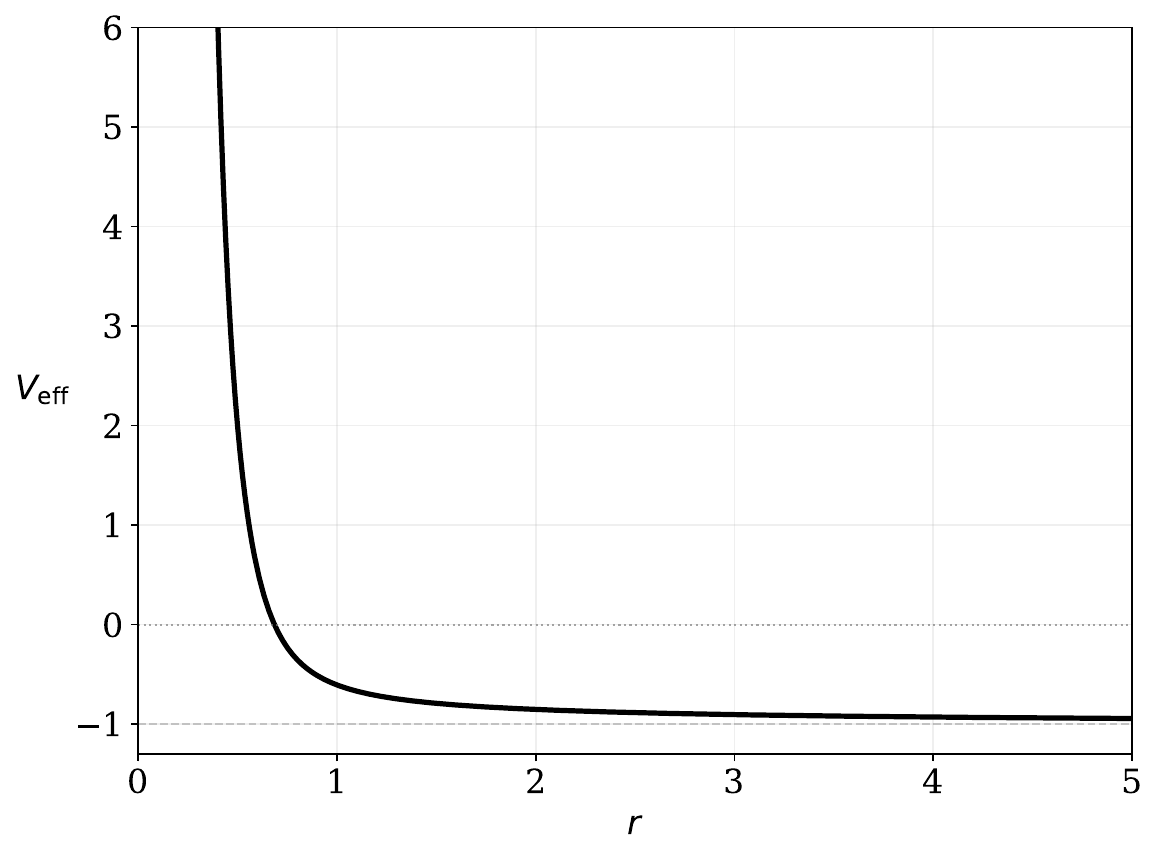}
\caption{\small A mixed-orientation solution of the quartic junction equation \eqref{eq:quartic} for three thermal-AdS pages ($\mu_i=0$) with $d=4$, $m=1$, $G=1$, 
$L_1=1$, $L_2=L_3=2$. The
physical branch has orientation $(-,+,+)$, approaches
$V_{\rm eff}\to-1$ at large radius, and crosses $V_{\rm eff}=0$ at the
turning point $r_*$.}
\label{fig:2}
\end{figure}

Figure~\ref{fig:2} illustrates a mixed-orientation thermal-AdS example with $d=4$, $L_1=1$, and $L_2=L_3=2$, for which Eq.~\eqref{eq:L-relation} is satisfied. The physical branch in the numerical example has orientation $(-,+,+)$. It approaches $V_{\rm eff}\rightarrow-1$ at large $r$, diverges positively at small $r$, and crosses zero at a turning point $r_*$. The mass parameters $\mu_i$ do not enter the leading $O(r)$ cancellation in Eq.~\eqref{eq:Or}; setting them to zero therefore isolates the unequal-$L_i$ mechanism. They can nevertheless affect the subleading behavior and the global existence of the trajectory.

Physically, the sign $\epsilon_i$ specifies the orientation of Euclidean-time flow on page $i$: $\epsilon_i=+1$ means that $\tau_i$ increases with the shell parameter $T$, whereas $\epsilon_i=-1$ means that it decreases. Thus, $(-,+,+)$ assigns the opposite Euclidean-time orientation to page 1 relative to pages 2 and 3. Whether such a page should be interpreted as a leg of the same state preparation or as part of a conjugate amplitude depends on the global sewing, which is not fixed by the local junction equation.

Unequal $L_i$ imply that the boundary CFTs dual to the three boundaries have different central charges --- in other words, this resembles entanglement between three distinct theories. Since the mass parameter $\mu_i$ enters $f_i$ only at ${\cal O}(r^{-(d-2)})$, it does not affect the leading condition \eqref{eq:Or}: the asymmetry is driven by the mixed orientation $\epsilon_i$ and the resulting unequal $L_i$, while a mass difference alone is a subleading effect. Gluing spacetimes with different $f_i$ through the Israel conditions is geometrically nontrivial and needs future investigation. In the following we will restrict our discussion to the case with equal $L_i$.

\subsection{Candidate profiles}
\label{sec:mixed-phases}

We next discuss the possible thermal-AdS and black-hole fillings of the three
pages.
In a factorized canonical ensemble, an individual page is associated with the large Euclidean black-hole saddle when $\beta_i<\beta_{\rm HP}$ and with thermal AdS when $\beta_i>\beta_{\rm HP}$ \cite{HawkingPage1983}. For the coupled booklet, these inequalities label candidate page fillings but do not by themselves establish a common saddle. A candidate trajectory must solve
\be
\dot r^2+V_{\rm eff}^{ijk}(r)=0,
\label{eq:Veff-ijk}
\ee
together with
\be
\begin{split}
&\epsilon_{i}\sqrt{f_i+V_{\rm eff}^{ijk}}
+\epsilon_{j}\sqrt{f_j+V_{\rm eff}^{ijk}}
+\epsilon_{k}\sqrt{f_k+V_{\rm eff}^{ijk}} = \kappa(r), \end{split}
\label{eq:Veff-ijk-def}
\ee
where
\be
\epsilon_{i,j,k}=\pm1,\quad \kappa(r)=\frac{8\pi G_{d+1}m}{(d-1)\Omega_{d-1}r^{d-2}}.
\ee
Equivalently, $V_{\rm eff}^{ijk}$ solves the quartic equation \eqref{eq:quartic} with $Y_n=f_n+V_{\rm eff}^{ijk}$. In addition, it must satisfy
\be
\dot r^2=-V_{\rm eff}^{ijk}(r)\ge0,
\qquad
r_*>0,
\qquad
V_{\rm eff}^{ijk}(r_*)=0,
\label{eq:saddle-cond}
\ee
as well as the Euclidean periodicity and regularity conditions on every page.

Table~\ref{tab:branches} summarizes the combinatorial page assignments and their status within the present local analysis, modulo permutations. Here we denote a profile by $X_{a|b|c}$, where $a,b,c\in\{\AdS,\BH\}$.

\begin{table}[h!]
\centering
\begin{tabular}{ll}
\toprule
Branch & Geometric interpretation \\
\midrule
$\AdS|\AdS|\AdS$ & three thermal-AdS pages (closed booklet cosmology) \\
$\BH|\BH|\BH$ & three-boundary black-hole booklet \\
$\AdS|\BH|\BH$ & one AdS leg and a two-BH booklet sector \\
$\AdS|\AdS|\BH$ & baby-universe-like sector and one BH leg \\
\bottomrule
\end{tabular}
\caption{\small Candidate page profiles allowed by the combinatorics of the
three-page construction.
The individual Hawking--Page criterion labels the candidate filling
on each page.}
\label{tab:branches}
\end{table}

\begin{figure}[h!]
\centering
\includegraphics[width=0.5\textwidth,trim=152pt 124pt 127pt 106pt,clip]{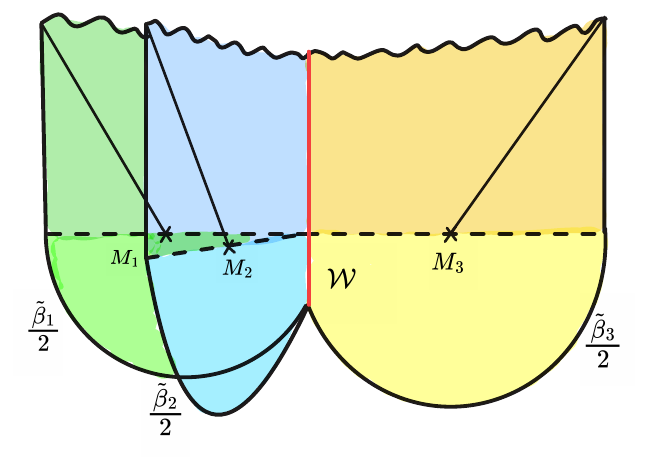}
\caption{\small Schematic of the $\BH|\BH|\BH$ branch: all three pages are black holes connected through the booklet shell worldvolume.}
\label{fig:3}
\end{figure}

Among the classes in Table~\ref{tab:branches}, two are especially relevant here. The first is the $\BH|\BH|\BH$ branch, shown schematically in Fig.\ \ref{fig:3}, which is the sector used in the state-counting analysis of Sec.~\ref{sec:counting}. Locally, it consists of three black-hole pages joined to the common dust-shell worldvolume. It is a candidate three-boundary booklet analogue of the two-boundary shell geometry. Referring to the three pages as sharing a common Lorentzian interior requires an additional global continuation.

\begin{figure}[h!]
\centering
\includegraphics[width=0.5\textwidth,trim=136pt 112pt 129pt 98pt,clip]{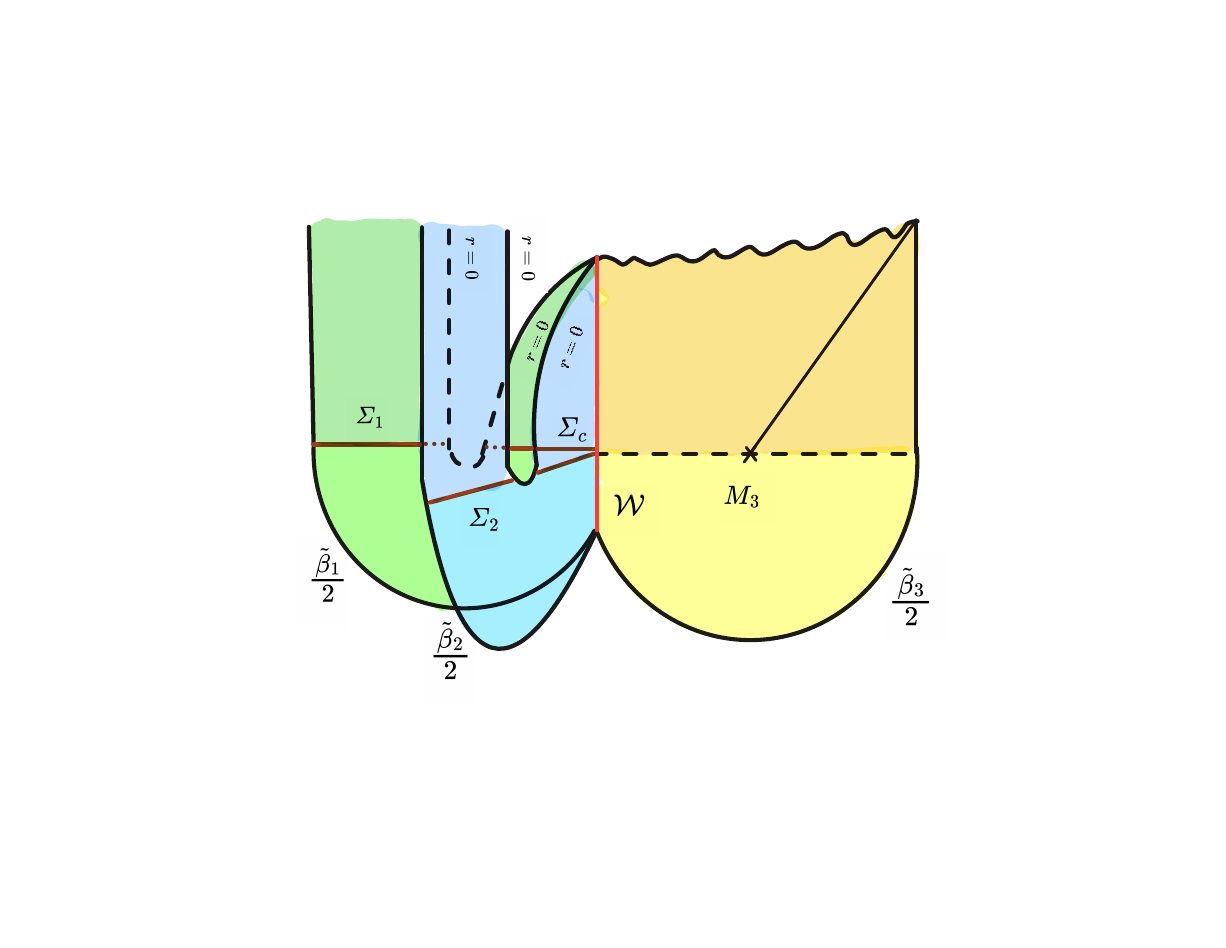}
\caption{\small Schematic of the $\AdS|\AdS|\BH$ branch.}
\label{fig:4}
\end{figure}

The second class is the $\AdS|\AdS|\BH$ candidate shown in Fig.\ \ref{fig:4}. Such a mixed filling may be realized with an admissible mixed orientation, and its two horizonless thermal-AdS pages could define a baby-universe-like sector \cite{Antonini2023,Antonini2025} coupled to a black-hole page.
This resembles the Type-B geometry in which a compact big-crunch sector is attached to a parent black-hole exterior \cite{BalasubramanianYildirim2025a}. Establishing this interpretation also requires a globally consistent Lorentzian continuation of the booklet saddle which we leave open.

In Sec.~\ref{sec:probes}, we study how heavy-operator probes respond to these configurations. Before that, we study the state counting using the symmetric all-BH branch.

\section{State Counting in the Symmetric All-BH Sector}
\label{sec:counting}

The junction data introduced in Sec.~\ref{sec:PETS3} can label arbitrarily large families of candidate $\pets_3$ states. The number of labels, however, need not equal the number of linearly independent quantum states. As in the two-boundary shell construction, nonperturbative Euclidean wormholes can generate small overlap moments and thereby reduce the rank of the associated Gram matrix \cite{Balasubramanian2024}. In this section, we extend that counting logic to three physical boundaries.

Throughout the section, we restrict attention to the symmetric all-BH branch identified in Sec.~\ref{sec:symmetric}: the three pages have the same AdS radius, the same black-hole mass, and the all-positive orientation.

\subsection{Single-shell states}
\label{sec:single-shell}

\begin{figure}[h!]
\centering
\includegraphics[width=0.55\textwidth,trim=110pt 116pt 126pt 108pt,clip]{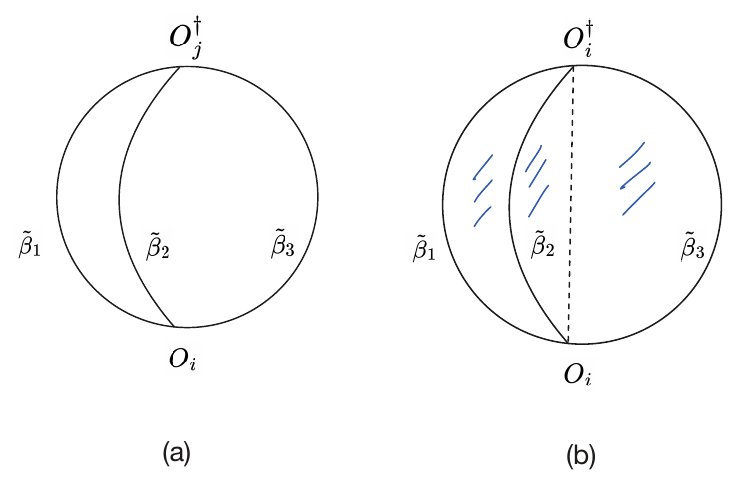}
\caption{\small Euclidean path integrals for overlaps of single-shell $\pets_3$ states, with the transverse $S^{d-1}$ suppressed. (a) Path integral for the overlap $\braket{\Psi_{m_{j}}}{\Psi_{m_{i}}}$ between two distinct $\pets_3$ states. (b) Overlap $\braket{\Psi_{m_{i}}}{\Psi_{m_{i}}}$ with identical shell operators. The dashed line is the trajectory of the dust shell in the dual bulk.}
\label{fig:5}
\end{figure}

We begin with states $\ket{\Psi_m}$ labeled by the inertial mass $m$ of a single junction shell.
The overlaps are computed by the Euclidean path integrals illustrated in Fig.~\ref{fig:5}.
From Fig.~\ref{fig:5}(b), we obtain $Z_1\simeq e^{-I_{\rm E}}$ which is the leading semiclassical normalization factor for the states $\ket{\Psi_{m_i}}$. Then we normalize the state with a factor $1/\sqrt{Z_1}$ for $\ket{\Psi_{m_i}}$ and explore the properties of these normalized states.

For two arbitrary shell labels, the leading connected contribution to the squared overlap is
\be
\avg{|\braket{\Psi_{m'}}{\Psi_{m}}|^2}
\simeq
\delta_{m,m'}
+
\frac{Z_2}{Z_1\,Z_{1'}},
\label{eq:overlap}
\ee
where $Z_1$ and $Z_{1'}$ are the single-state normalization factors and $Z_2\simeq e^{-I_{\rm wh}}$ is the connected wormhole contribution. We use $\avg{\cdots}$ to denote the coarse-grained gravitational average obtained from the relevant semiclassical saddles.

\begin{figure}[h!]
\centering
\includegraphics[width=0.78\textwidth,trim=70pt 171pt 75pt 121pt,clip]{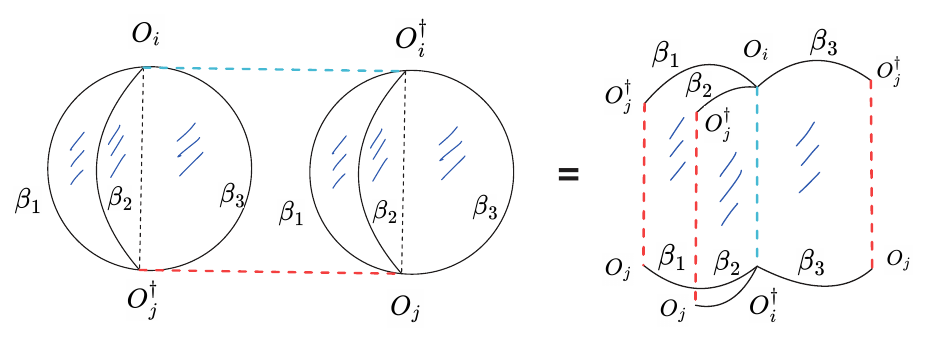}
\caption{\small Overlap $\avg{|\braket{\Psi_{m'}}{\Psi_m}|^2}$ in the heavy-shell limit. Each of the three physical page
sectors contributes a cylinder factor $Z(2\beta)$, giving
\eqref{eq:overlap2} after normalization.}
\label{fig:6}
\end{figure}

In the heavy-shell limit $m,m'\gg M$, where $M$ is the common black-hole mass,
the Euclidean propagation time of each shell tends to zero. Consequently,
$\tilde\beta_m\approx\tilde\beta_{m'}\approx\beta$, and the shell action becomes a constant independent of the exterior black-hole data at leading order. The connected geometry then reduces to three independent cylinders of circumference $2\beta$, one for each physical page, as shown in Fig.~\ref{fig:6}. For distinct states,
\be
\avg{|\braket{\Psi_{m'}}{\Psi_{m}}|^2}
\simeq
\frac{Z(2\beta)^3}{Z(\beta)^6},
\label{eq:overlap2}
\ee
where $Z(\beta)$ is the gravitational partition function with Euclidean period $\beta$.

\begin{figure}[h!]
\centering
\includegraphics[width=0.45\textwidth,trim=174pt 140pt 162pt 100pt,clip]{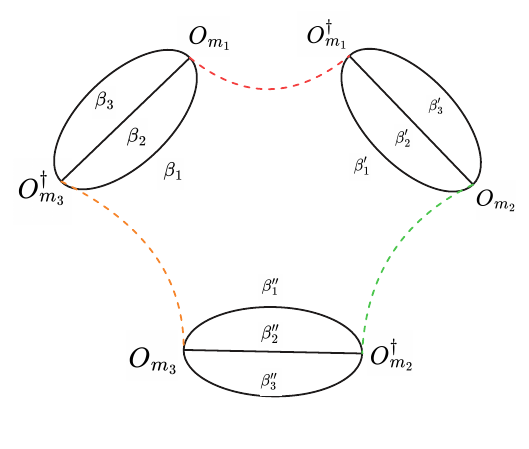}
\caption{\small Maximally connected three-boundary wormhole contributing to the connected part of the third overlap moment. Thin shells propagate along the dashed colored lines; the black holes are the boundaries of the geometry.}
\label{fig:7}
\end{figure}

We next consider the connected part of the $n$-th cyclic overlap moment. We assume that its semiclassical contribution is the maximally connected replica wormhole; Fig.~\ref{fig:7} illustrates the case $n=3$. In the heavy-shell limit, this assumption gives
\be
\avg{\braket{\Psi_{m_1}}{\Psi_{m_2}}\cdots\braket{\Psi_{m_n}}{\Psi_{m_1}}}_c
\simeq
\frac{Z(n\beta)^3}{Z(\beta)^{3n}}.
\label{eq:nth-moment}
\ee
Here the subscript $c$ denotes the connected part. The cube again counts the three physical page sectors; $n$ is the replica order and should not be confused with the number of asymptotic boundaries.

\subsection{Multi-shell states}
\label{sec:multi-shell}

We now consider a state $\ket{\Psi_{\mathbf m}}$ with shell data $\mathbf m=(m_1,\ldots,m_k)$. If the interior is constructed exclusively from trivalent $Y$-shaped $\pets_3$ junctions and has three external legs, the number $k$ of internal junction vertices must be odd. Indeed, a trivalent graph with $k$ internal vertices, $E_{\rm int}$ internal edges, and three external legs satisfies
\[
3k=2E_{\rm int}+3,
\qquad
E_{\rm int}=\frac{3(k-1)}{2}.
\]
The possible gluing patterns can be related to connected cubic graphs obtained by adjoining one additional trivalent vertex to the three external legs. One may enlarge the construction by allowing two-leg $\pets_2$ components, in which case arbitrary values of $k$ become possible. The formulas below, however, refer to the purely trivalent sector; an enlarged graph family requires the corresponding internal-edge bookkeeping.

In the heavy-shell limit, the normalization of a trivalent $k$-junction state has the schematic form
\be
Z_1^{(\mathbf m)}
\simeq
Z(\beta)^3 e^{-\sum_{n=1}^{k}I_{n}}
\prod_{i=1}^{\frac{3(k-1)}{2}} Z(\beta_i),
\label{eq:Z1-multi}
\ee
where $k$ is the number of junction shells. The three factors of $Z(\beta)$ come from the exterior legs, while each $Z(\beta_i)$ is associated with an internal Euclidean segment. Here $I_i$ is the action of the $i$th shell. The shell actions add because all shell propagation times vanish at leading order in the heavy-shell limit.

\begin{figure}[h!]
\centering
\includegraphics[width=0.85\textwidth,trim=141pt 92pt 115pt 50pt,clip]{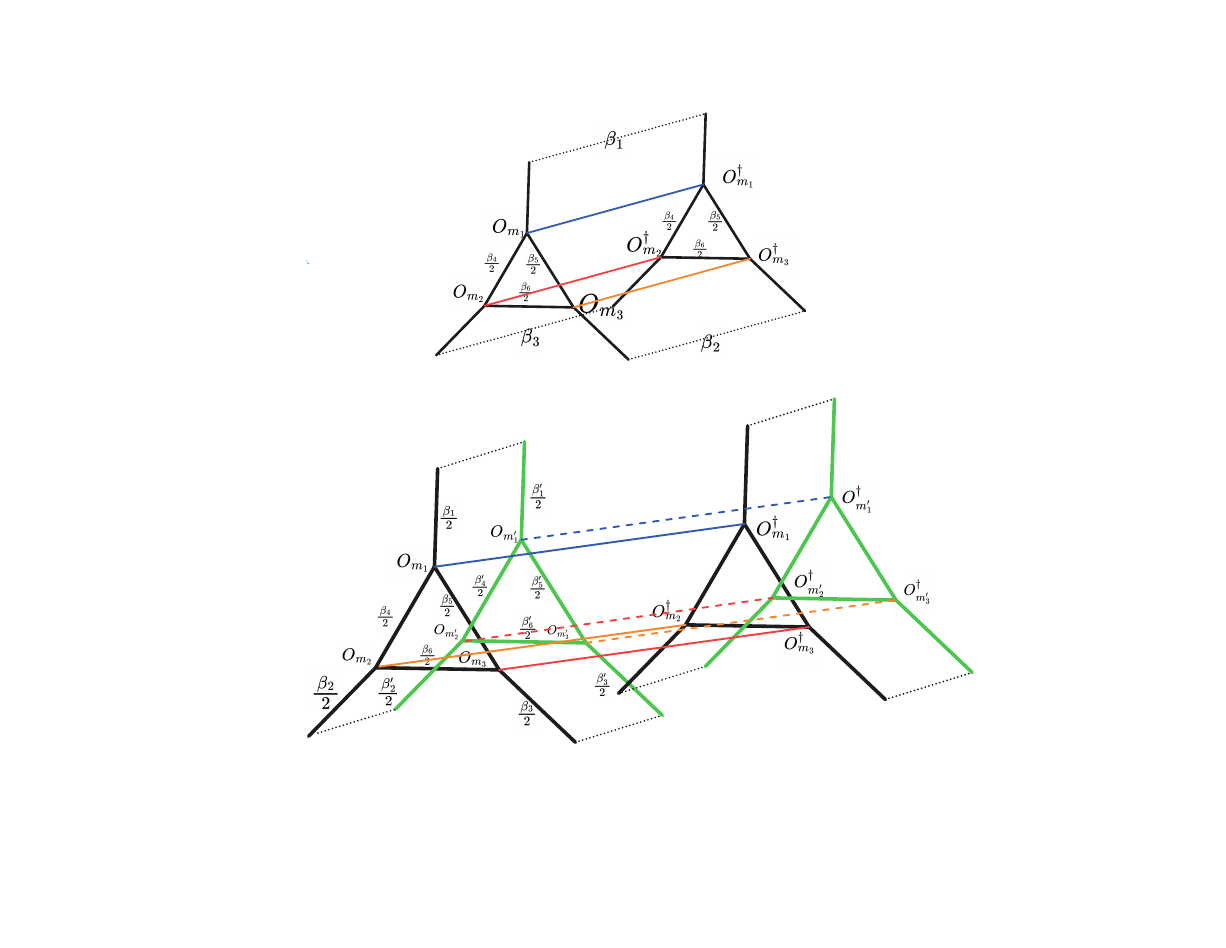}
\caption{\small Example of a 3-shell state. The upper diagram shows the normalization in the denominator, $Z_1^{(\mathbf m)}$, generalizing the normalization of the single-shell state in Fig.~\ref{fig:5}(b). For a 3-shell state, this normalization factor is given by $Z(\beta)^3 Z(\beta/2)^3$. The lower diagram corresponds to the wormhole contribution of the second moment, $Z_2$, generalizing the single-shell case in Fig.~\ref{fig:6}. Calculation yields this part to be $Z(2\beta)^3 Z(\beta/2)^6$. Overall, the overlap of the 3-shell state in Eq.\ \eqref{eq:overlap-universal} is given by $\frac{Z(2\beta)^3 Z(\beta/2)^6}{(Z(\beta)^3 Z(\beta/2)^3)^2} = \frac{Z(2\beta)^3}{Z(\beta)^6}$.}
\label{fig:8}
\end{figure}

For the connected second moment between states with $k$ and $k'$ junctions, the corresponding numerator is
\be
Z_2^{(\mathbf m,\mathbf m')} \simeq Z(2\beta)^3 \, e^{-\sum_{n=1}^k I_{n} - \sum_{m=1}^{k'}I_{m}} \, \prod_{i=1}^{\frac{3(k-1)}{2}} Z(\beta_i) \prod_{j=1}^{\frac{3(k'-1)}{2}} Z(\beta_j').
\label{eq:Z2-multi}
\ee
The three factors of $Z(2\beta)$ come from the exterior legs. The factors $Z(\beta_i)$ and $Z(\beta_j')$ arise from internal segments. Fig.~\ref{fig:8} shows an example of a three-shell state.

Therefore, the connected overlap reduces to the same universal quantity as in the single-shell case:
\be
\avg{|\braket{\Psi_{\mathbf m'}}{\Psi_{{\mathbf m}}}|^2}_c
\simeq
\frac{Z(2\beta)^3}{Z(\beta)^6}.
\label{eq:overlap-universal}
\ee

In the heavy-shell limit, with all shell masses much larger than $M$, the same cancellation can extend to the $n$-th moment of the multi-shell states:
\be
\avg{\braket{\Psi_{\mathbf{m}_1}}{\Psi_{\mathbf{m}_2}}\braket{\Psi_{\mathbf{m}_2}}{\Psi_{\mathbf{m}_3}}\cdots\braket{\Psi_{\mathbf{m}_n}}{\Psi_{\mathbf{m}_1}}}_c
=
\frac{Z_n}{Z_1^{(\mathbf{m}_1)}\cdots Z_1^{(\mathbf{m}_n)}}\simeq
\frac{Z(n\beta)^3}{Z(\beta)^{3n}}.
\label{eq:nth-moment-multi}
\ee
This result will be used in the Gram-matrix analysis below.

\subsection{Gram matrix and Hilbert space dimension}
\label{sec:gram}

We now show that, similarly to the $\pets_2$ case, the number of $\pets_3$ states does not overcount the Bekenstein-Hawking entropy. Consider $K$ candidate three-boundary states drawn from the shell family and define their
Gram matrix by
\be
G_{ab}^{(3)}
\equiv
\braket{\Psi_a^{(3)}}{\Psi_b^{(3)}},
\qquad a,b = 1,\ldots,K.
\label{eq:Gram3}
\ee
Writing $G^{(3)}=B^\dagger B$, where the columns of $B$ are the state vectors, shows that $G^{(3)}$ is Hermitian and positive semidefinite. Its rank is the number of linearly independent states in the family.

We extract the rank of the Gram matrix from the resolvent, which is defined as
\be
R_{ij}(\lambda)
\equiv
\left(\frac{1}{\lambda\,\mathbb{I}-G^{(3)}}\right)_{ij}
=
\frac{1}{\lambda}\,\delta_{ij}
+
\sum_{n=1}^{\infty}\frac{1}{\lambda^{n+1}}\,(G^{n})_{ij}.
\label{eq:resolvent}
\ee
Its trace $R(\lambda)=\sum_{i=1}^{K}R_{ii}(\lambda)$ determines the averaged Gram-matrix eigenvalue density through
\be
D(\lambda)
=
\frac{1}{2\pi i}\left(R(\lambda-i\epsilon)-R(\lambda+i\epsilon)\right).
\label{eq:spectral-density}
\ee
The moments $(G^{n})_{ij}$ are products of $n$ overlaps. From Secs.\ \ref{sec:single-shell} and \ref{sec:multi-shell}, the connected part of a cyclic product of $n$ overlaps \eqref{eq:nth-moment-multi} is independent of the shell masses. In the leading semiclassical approximation, only planar diagrams contribute, and the resolvent is determined self-consistently by a Schwinger-Dyson equation \cite{Penington2022}. To use the canonical moments in \eqref{eq:nth-moment-multi}, we project each of the three physical boundaries into a microcanonical window, since each Gram-matrix entry is a sum over energy bands. The three-boundary canonical partition function is
\be
Z(\beta)^3
=
\int dE_1 dE_2 dE_3\,
e^{-\beta(E_1+E_2+E_3)}
e^{S(E_1)+S(E_2)+S(E_3)},
\label{eq:Z3-can}
\ee
where the cube reflects the three independent pages. Projecting each boundary energy independently onto the same narrow microcanonical window $[E-\Delta E/2,E+\Delta E/2]$ gives:

\be
d(E)
=
e^{S_1(E)+S_2(E)+S_3(E)}
=
e^{3S(E)}.
\label{eq:D3}
\ee
At leading exponential order, we identify $z(E)\Delta E\simeq e^{S(E)}$. Define the three-boundary microcanonical quantities
\be
e^{\mathbf{S}_3} \equiv (z(E)\,\Delta E)^3,
\qquad
\mathbf{Z}_n \equiv (z(E)\, e^{-n\beta E}\,\Delta E)^3,
\label{eq:micro-can}
\ee
so that the ratio entering the resolvent equation is
\be
\frac{\mathbf{Z}_n}{(\mathbf{Z}_1)^n}
=
\frac{(z\,e^{-n\beta E}\Delta E)^3}{(z\,e^{-\beta E}\Delta E)^{3n}}
=
e^{-3(n-1)S(E)}
\equiv
e^{-(n-1)\mathbf{S}_3}.
\label{eq:ratio}
\ee
The Schwinger-Dyson equation for the trace of the resolvent then reads
\be
\lambda\,\avg{R(\lambda)}
=
K
+
\sum_{n=1}^{\infty}\frac{\mathbf{Z}_n}{(\mathbf{Z}_1)^n}\,\avg{R(\lambda)}^{\,n}
=
K
+
\frac{e^{\mathbf{S}_3}\,\avg{R(\lambda)}}{e^{\mathbf{S}_3}-\avg{R(\lambda)}},
\label{eq:SD}
\ee
which rearranges into the quadratic equation
\be
\avg{R(\lambda)}^{\,2}
+
\left(\frac{e^{\mathbf{S}_3}-K}{\lambda}-e^{\mathbf{S}_3}\right)\avg{R(\lambda)}
+
\frac{K}{\lambda}\,e^{\mathbf{S}_3}
=
0.
\label{eq:quadratic}
\ee

The resulting eigenvalue density has the Marchenko-Pastur form,
\be
\begin{split}
\avg{D(\lambda)}
&=
\frac{e^{\mathbf{S}_3}}{2\pi\lambda}
\sqrt{\left[\lambda-\left(1-K^{1/2}e^{-\mathbf{S}_3/2}\right)^{2}\right]
\left[\left(1+K^{1/2}e^{-\mathbf{S}_3/2}\right)^{2}-\lambda\right]}
\\
&\quad+
\delta(\lambda)\left(K-e^{\mathbf{S}_3}\right)\theta\!\left(K-e^{\mathbf{S}_3}\right).
\end{split}
\label{eq:MP}
\ee
The $\delta$-function contribution counts the zero eigenvalues when $K > e^{\mathbf{S}_3}$. The rank is therefore
\be
\operatorname{rank} G^{(3)}
=
\min\bigl(K, e^{\mathbf{S}_3}\bigr).
\label{eq:rank}
\ee
For $K\geq e^{\mathbf{S}_3}$, the rank saturates at the dimension of the three-boundary microcanonical product space $e^{\mathbf{S}_3}$.

\section{Heavy-Operator Probes of Three-Boundary Booklet States}
\label{sec:probes}

Fine-grained preparation data may be invisible to simple boundary observables.
Nevertheless, a fine-tuned heavy-shell probe can test a proposed shell state even when the relevant information is hidden behind a horizon \cite{BalasubramanianYildirim2025a}.
For a matched probe, the path integral contains additional detection wormhole saddles, producing an order-one enhancement in the one- and two-boundary constructions. We extend this Euclidean diagnostic to three-boundary states using the nonperturbative framework of Refs.~\cite{BalasubramanianYildirim2025a,BalasubramanianYildirim2025b}.

We focus on the symmetric all-BH branch of Sec.~\ref{sec:symmetric}. The value of the detection ratio depends on the saddle enumeration and on a factor-counting approximation to the multivariable inverse Laplace transform. Its interpretation as a Lorentzian measurement would additionally require a consistent analytic continuation.

\subsection{Probe Setup}
\label{sec:probe-setup}

For a $\pets_3$ state $\ket{\Psi_{m_i}}$, a heavy spherically symmetric shell operator $O_P$ of inertial mass $m_P\sim O(1/G_N)$ is inserted on boundary 1. On $\Hcal_1\otimes\Hcal_2\otimes\Hcal_3$, a probe localized on boundary 1 acts globally as $O_P^{(1)}\otimes\mathbb{I}_2\otimes\mathbb{I}_3$. We denote this tensor-product operator simply by $O_P$. Under the assumed random-phase ensemble of Sec.~\ref{sec:GHZ}, the one-point function $\avg{\bra{\Psi_{m_i}}O_P\ket{\Psi_{m_i}}}$ vanishes after coarse-graining, while the squared matrix element $\avg{\bigl|\bra{\Psi_{m_{i}}}O_P\ket{\Psi_{m_{i}}}\bigr|^2}$ can receive nonzero wormhole contributions.
This is computed by a Euclidean gravitational path integral which receives contributions from two classes of saddle-point geometries: for an unmatched probe, only universal saddles contribute; for a matched probe, detection saddles also contribute. Here and below, we use $P\sim i$ to denote the ``match", meaning $m_P=m_i$ and agreement of all other shell quantum numbers relevant to the semiclassical saddle.

$Z_U$ denotes the universal contribution present for both matched and unmatched probes, and $Z_D$ denotes the additional contribution present only for a matched probe. Normalizing the unmatched response to unity, the detection-to-universal ratio is
\be
\frac{Z_U+Z_D}{Z_U}=1+\frac{Z_D}{Z_U}.
\label{eq:probe-response}
\ee

\subsubsection{Universal saddles}

\begin{figure}[h!]
\centering
\includegraphics[width=0.85\textwidth,trim=66pt 38pt 52pt 37pt,clip]{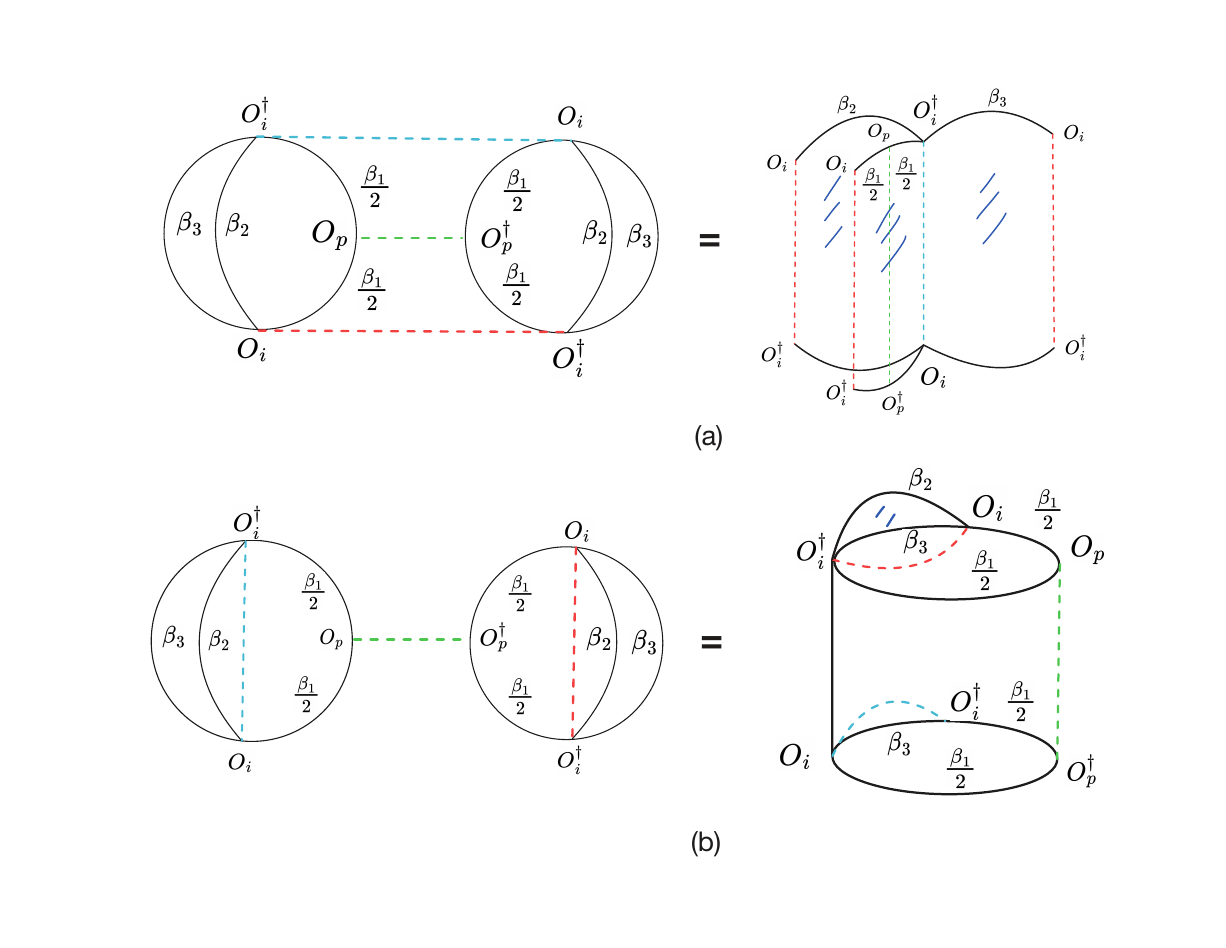}
\caption{\small The two universal saddle classes contributing to $\avg{\bra{i}O_P\ket{i}\bra{i}O_P^\dagger\ket{i}}$ for an unmatched probe, $P\nsim i$. The left panels show the shell asymptotic boundary conditions, and the right panels show the corresponding bulk shell trajectories.}
\label{fig:9}
\end{figure}

For the universal saddles, the probe shells propagate and annihilate with each other. There are two classes of cylinder-topology saddles that contribute universally, shown in Fig.~\ref{fig:9}. Here we denote the three-boundary state with shell mass $m_i$ by the state $\ket{i}$; for the universal saddles it is not necessary that the shell mass $m_i$ equal the probe mass $m_P$.

In the first class, the shells propagate between the two boundaries of the cylinder. The probe shells annihilate on boundary 1, while the preparation shells associated with boundaries 2 and 3 propagate across the full cylinder, see Fig.~\ref{fig:9}(a). In the large-shell-mass limit, the shell propagation times on the disks vanish, so the disk circumferences are $\beta_1$, $2\beta_2$, and $2\beta_3$. We use $\bar Z(\beta)$ to denote the black-hole-sector gravitational partition function with Euclidean period $\beta$. The corresponding sheet diagram yields
\be
\bar Z(\beta_1)^2 \, \bar Z(2\beta_2) \, \bar Z(2\beta_3).
\label{eq:ZU1}
\ee
In the second class, the probe shell propagates between the two sides of the cylinder, while the preparation shells associated with boundary 1 propagate across it, see Fig.~\ref{fig:9}(b). The corresponding contribution is
\be
\bar Z(\beta_2)^2 \, \bar Z(\beta_3)^2 \, \bar Z(2\beta_1).
\label{eq:ZU2}
\ee

Each preparation shell contributes a matter factor $Z_{m_i}\sim e^{-I_{\rm shell}(m_i)}$, and the probe contributes $Z_{m_P}\sim e^{-I_{\rm shell}(m_P)}$. Their precise normalization is not needed because the common product $Z_{m_i}^2Z_{m_P}$ cancels from the final ratio. The universal contribution is therefore
\be
Z_U
=
\bigl(\bar Z(\beta_1)^2 \, \bar Z(2\beta_2) \, \bar Z(2\beta_3)
+
\bar Z(\beta_2)^2 \, \bar Z(\beta_3)^2 \, \bar Z(2\beta_{1})\bigr) \times Z_{m_i}^2 Z_{m_P}.
\label{eq:ZU}
\ee

\subsubsection{Detection saddles}

\begin{figure}[h!]
\centering
\includegraphics[width=0.85\textwidth,trim=84pt 14pt 72pt 0pt,clip]{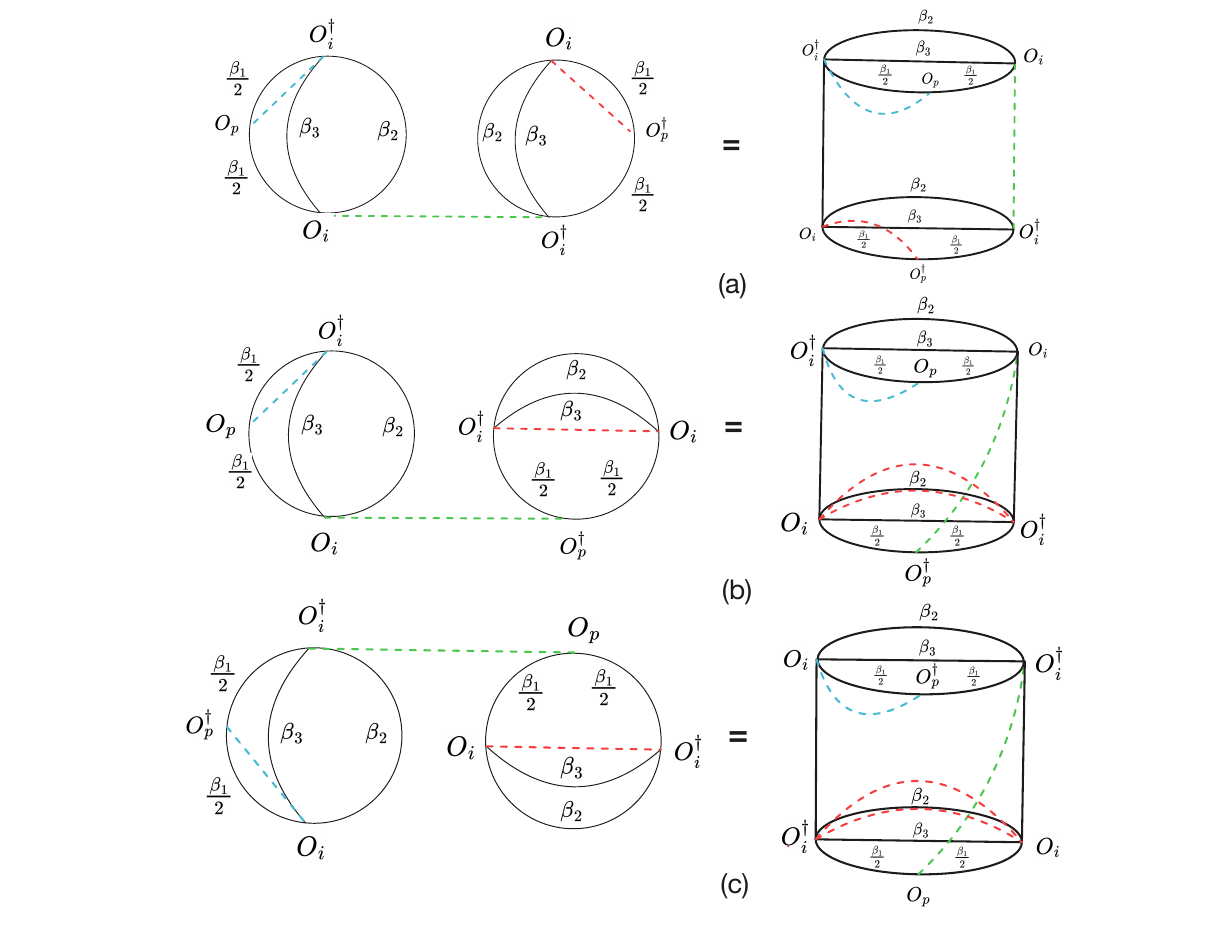}
\caption{\small The first three detection saddles for a matched probe, $P\sim i$. The second and third saddles are symmetric under $O_{i}\leftrightarrow O_{i}^\dagger$, producing the factor of two in Eq.\ \eqref{eq:ZD2}.}
\label{fig:10}
\end{figure}

\begin{figure}[h!]
\centering
\includegraphics[width=0.6\textwidth,trim=140pt 168pt 132pt 103pt,clip]{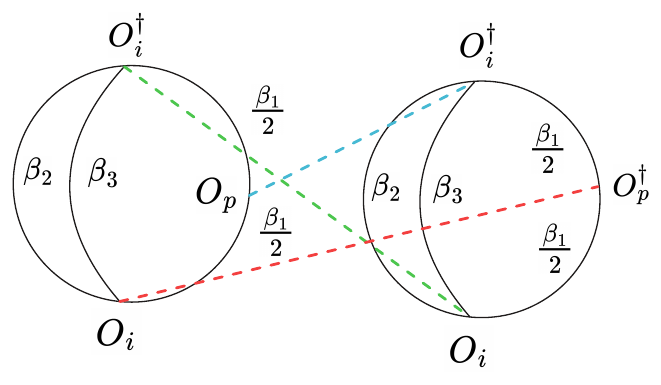}
\vspace{-0.2cm}
\caption{\small The fourth detection saddle class, represented by a
nonintersecting routing on a twice-punctured torus.}
\label{fig:11}
\end{figure}

When $P\sim i$, a probe shell may instead annihilate with a preparation shell. The sheet construction then admits the four additional saddle classes shown in Figs.~\ref{fig:10} and \ref{fig:11}. The first three are variations on the universal saddles, differing only in which shells are paired. For class 1, the probe shell on boundary 1 annihilates with the preparation shell on the same boundary in one of the four strip copies. The remaining preparation shells, originally inserted on boundaries 2 and 3, propagate across the resulting connected manifold. In the large-shell-mass limit, all propagation times pinch off and the shell homology regions shrink to zero, yielding the factor
\be
D_{1}=\bar Z(2\beta_1 + \beta_2) \, \bar Z(2\beta_3 + \beta_1) \, \bar Z(\beta_1/2)^2 \, \bar Z(2\beta_2 + \beta_3) \times Z_{m_{i}}^2 Z_{m_P}.
\label{eq:ZD1}
\ee
Classes 2 and 3 give equal contributions because they are exchanged by $O_i\leftrightarrow O_i^\dagger$. In the large-shell-mass limit,
\be
D_{2} + D_{3}=2 \, \bar Z(\beta_2) \, \bar Z(\beta_3) \, \bar Z(\beta_1/2) \, \bar Z\!\left(\frac{3}{2}\beta_1 + 2\beta_2\right) \, \bar Z\!\left(\frac{3}{2}\beta_1 + 2\beta_3\right)\times Z_{m_{i}}^2 Z_{m_P}.
\label{eq:ZD2}
\ee

The fourth saddle, shown in Fig.~\ref{fig:11}, has a distinct topology. In the proposed sheet representation, the shell trajectories form a nonintersecting Eulerian routing on a twice-punctured torus. In the large-shell-mass limit, its contribution is $D_4=\bar Z(2\beta_1+2\beta_2+2\beta_3)Z_{m_i}^2Z_{m_P}$.

Adding the four classes, we obtain
\begin{equation}
\begin{split}
Z_D
&= \bigl(\bar Z(2\beta_1 + \beta_2) \, \bar Z(2\beta_3 + \beta_1) \, \bar Z(\beta_1/2)^2 \, \bar Z(2\beta_2 + \beta_3) \\
&\quad + 2 \, \bar Z(\beta_2) \, \bar Z(\beta_3) \, \bar Z(\beta_1/2) \,
\bar Z\!\left(\frac{3}{2}\beta_1 + 2\beta_2\right) \,
\bar Z\!\left(\frac{3}{2}\beta_1 + 2\beta_3\right) \\
&\quad + \bar Z(2\beta_1 + 2\beta_{2}+2\beta_{3})\bigr)\times Z_{m_{i}}^2 Z_{m_P}.
\label{eq:ZD}
\end{split}
\end{equation}
The common shell factors cancel between $Z_D$ and $Z_U$, leaving the detection-to-universal ratio
\begin{equation}
\begin{split}
\frac{Z_D}{Z_U}
&=
\Big[
\bar Z(2\beta_1+\beta_2)
\bar Z(2\beta_3+\beta_1)
\bar Z(\tfrac{\beta_1}{2})^2
\bar Z(2\beta_2+\beta_3)
\\
&\qquad
+2\bar Z(\beta_2)\bar Z(\beta_3)\bar Z(\tfrac{\beta_1}{2})
\bar Z(\tfrac32\beta_1+2\beta_2)
\bar Z(\tfrac32\beta_1+2\beta_3)
+\bar Z(2\beta_1 + 2\beta_{2}+2\beta_{3})
\Big]
\\
&\qquad\times
\Big[
\bar Z(\beta_1)^2 \bar Z(2\beta_2)\bar Z(2\beta_3)
+
\bar Z(\beta_2)^2\bar Z(\beta_3)^2\bar Z(2\beta_1)
\Big]^{-1}.
\end{split}
\label{eq:ratio-full}
\end{equation}

\subsection{Detection ratio}
\label{sec:detection-ratio}

To evaluate the ratio \eqref{eq:ratio-full}, we project the external shell states into microcanonical windows using the same inverse-Laplace prescription as in the two-boundary calculation \cite{BalasubramanianYildirim2025a}. Denote the canonical shell state with preparation inverse temperatures $\boldsymbol\beta=(\beta_1,\beta_2,\beta_3)$ by $\ket{i;\boldsymbol\beta}$. Its projection onto a window $[E_a-\Delta_a/2,E_a+\Delta_a/2]$ on each boundary is schematically
\be
\ket{i;\boldsymbol\beta,\mathbf E}
=
\prod_{a=1}^{3}
\left[
\int_{E_a-\frac{\Delta_a}{2}}^{E_a+\frac{\Delta_a}{2}}
d\widetilde E_a\,e^{-\frac{\beta_a}{2}\widetilde E_a}
\int_{\mathcal C_a}\frac{d\widetilde\beta_a}{2\pi i}\, e^{\frac{\widetilde E_a\widetilde\beta_a}{2}}
\right]
\ket{i;\widetilde{\boldsymbol\beta}} .
\label{eq:laplace}
\ee
Here $\mathcal C_a$ is the appropriate Bromwich contour.
The inner inverse transform projects onto fixed energy, while the outer integral restricts the result to the chosen window.
The squared matrix element contains four external shell states. Its full microcanonical projection therefore requires four independent copies of \eqref{eq:laplace}, as in the two-boundary construction \cite{BalasubramanianYildirim2025a}, and hence involves twelve independent inverse-temperature variables $\widetilde\beta_a^{(r)}$ with $a=1,2,3$ and $r=1,\ldots,4$. These variables must remain independent until all inverse transforms have been performed. Only then may the corresponding energy windows be identified. This point is important because the mixed arguments of the partition functions impose coupled energy constraints.

For each canonical partition-function factor, write
\[
\bar Z(x)
=
\int dE'\,e^{S_{\rm BH}(E')-xE'}.
\]
Its inverse Laplace transform is evaluated in the saddle-point approximation \cite{BalasubramanianYildirim2025a},
\be
\int dx\,e^{Ex}\bar Z(x)
\simeq
\sqrt{\frac{2\pi}{h}}\,e^{S_{\rm BH}(E)},
\qquad
e^{\mathbf S(E)}
\equiv
\Delta E\sqrt{\frac{2\pi}{h}}\,e^{S_{\rm BH}(E)},
\label{eq:saddle-transform}
\ee
where $h$ is the Hessian at the canonical saddle and $e^{\mathbf S(E)}$ is the degeneracy in the narrow microcanonical window. For an isolated factor with argument $x=\sum_a c_a\beta_a$, the saddle is associated schematically with the weighted energy
\[
E_{\rm eff}
=
\frac{\sum_a c_aE_a}{\sum_a c_a}.
\]

We now specialize to symmetric all-BH windows,
\[
E_1=E_2=E_3\equiv E,
\qquad
\Delta_1=\Delta_2=\Delta_3\equiv\Delta,
\]
and write $\mathbf S(E)\equiv S$. For a single mixed factor, equal windows give $E_{\rm eff}=E$; for example, $\bar Z(2\beta_1+\beta_2)$ is associated with $\mathbf S[(2E_1+E_2)/3]=S$. The following estimate assumes that, after restoring the four replica labels and carrying out the full twelve-variable inverse transform, the leading saddle weights reduce to the factor-counting prescription encoded in \eqref{eq:ZU}--\eqref{eq:ZD}. Under this assumption, and after removing common Boltzmann and shell-mass factors, each independent black-hole partition-function factor contributes one factor of $e^S$:
\[
Z_U\simeq e^{4S}+e^{5S}=e^{5S}(1+e^{-S}),
\]
and
\[
Z_D\simeq 3e^{5S}+e^S=e^{5S}(3+e^{-4S}).
\]
The all-BH detection-to-universal ratio is consequently
\be
\frac{Z_D}{Z_U}
\simeq
\frac{3+e^{-4S}}{1+e^{-S}}
\simeq3,
\qquad S\gg1.
\label{eq:ratio-final}
\ee
Thus, within this prescription, $Z_D/Z_U\simeq3$ and the full matched-to-unmatched response is $1+Z_D/Z_U\simeq4$.

For the mixed branches, when some boundaries sit below the black-hole threshold, the corresponding $S(E_i)$ vanishes on the thermal-AdS legs, and only the probe black-hole leg contributes the entropy factor $e^S$. The dominant detection and universal sectors contain an equal number of black-hole factors. Consequently, the powers of $e^S$ cancel out, leaving only an ${\cal O}(1)$ numerical coefficient.

\begin{table}[h!]
\centering
\begin{tabular}{lccc}
\toprule
& 1-boundary state & 2-boundary state & 3-boundary state \\
\midrule
$Z_D/Z_U$ (all BH) & $\simeq 1$ & $3/2$ & $3$ \\
\bottomrule
\end{tabular}
\caption{\small Detection-to-universal ratios under the corresponding saddle and
factor-counting prescriptions.}
\label{tab:comparison}
\end{table}

We may use the probes to distinguish different boundary numbers. The detection ratio of the three-boundary state is twice that of the two-boundary state, which in principle can be used to distinguish $n=2$ and $n=3$ geometries from one another.

\section{Conclusion and Discussion}
\label{sec:conclusion}

In this work, we developed a conjectural three-boundary extension of the PETS framework. We used the thermal GHZ state as an undeformed reference state and replaced its diagonal tensor with general three-index junction amplitudes dressed by independent Euclidean filters. An ETH-inspired statistical ansatz for these amplitudes provides a simple description of coarse-grained one-boundary thermality: within the dominant energy window, each reduced state takes an approximately Gibbs form, while the full three-boundary state remains pure.

On the bulk side, we formulated a booklet-inspired Euclidean preparation and derived the corresponding local multiway Israel conditions. The symmetric, all-positive solution gives the all-BH branch used in the subsequent state-counting analysis. Mixed orientation signs allow cancellations among the junction terms and can support pages with unequal AdS radii. Note that distinct AdS radii correspond to different CFT central charges from the boundary perspective, complicating a unified geometric interpretation of the common shell area. This asymmetry is a genuinely new feature of the three-page junction.

We next used connected wormhole moments to analyze the Gram matrix of the shell-state family. In the symmetric all-BH sector, the rank of a generic three-index family saturates at the dimension of the three-boundary microcanonical product space, $e^{{\bf S}_3}$, once the number of candidate states is sufficiently large.

Finally, we extended the heavy-shell probe diagnostic to one-boundary measurements of the proposed three-boundary states. We identified universal saddles that contribute for any probe and additional detection saddles that
appear when the probe matches the preparation data.
In particular, the all-BH three-boundary configuration is estimated to give $Z_D/Z_U \simeq 3$, which differs from the two-boundary all-BH ratio of $3/2$. This indicates that, in principle, a single-boundary observer equipped with heavy shell operators may determine the number of entangled boundaries purely from the probe response, but the signal is only an ${\cal O}(1)$ effect.

Further work would primarily sharpen the microscopic and Lorentzian interpretation of this picture. In particular, an explicit global sewing of the booklet pages would clarify matter propagation across the common interface, while the full replica-labeled inverse Laplace transform would determine the relative prefactors in the detection ratio. Related directions include mixed-phase solutions and more general junction tensors, such as cyclic and W-type sectors. Lorentzian and quantum matter-junction conditions provide useful tools for these extensions \cite{JiangLiu2026}, while large-$N$ filtering constructions offer a complementary perspective on Euclidean wormholes, closed universes, and black-hole interiors \cite{Liu2025,Penington2022}. More broadly, booklet geometries provide a promising setting in which to explore multipartite entanglement beyond the class of smooth single-manifold bulk duals.

\section*{Acknowledgements}
\enlargethispage{\baselineskip}

We thank Libo Jiang, Cheng Peng, and Ya-Wen Sun for useful discussions. This work is supported by the National Natural Science Foundation of China under grants Nos. 12375041 and 12575046.


\begin{thebibliography}{99}
\setlength{\itemsep}{5pt}
\setlength{\parskip}{0pt}

\bibitem{Bekenstein1973}
J.~D.~Bekenstein,
{\em Black Holes and Entropy,}
\href{https://doi.org/10.1103/PhysRevD.7.2333}{Phys. Rev. D \textbf{7}, 2333 (1973)}.

\bibitem{Hawking1975}
S.~W.~Hawking,
{\em Particle Creation by Black Holes,}
\href{https://doi.org/10.1007/BF02345020}{Commun. Math. Phys. \textbf{43}, 199 (1975)}.

\bibitem{Balasubramanian2024}
V.~Balasubramanian, A.~Lawrence, J.~M.~Magan, and M.~Sasieta,
{\em Microscopic Origin of the Entropy of Black Holes in General Relativity,}
\href{https://doi.org/10.1103/PhysRevX.14.011024}{Phys. Rev. X \textbf{14}, 011024 (2024)}
\href{https://arxiv.org/abs/2212.02447}{[arXiv:2212.02447 [hep-th]]}.

\bibitem{Balasubramanian2024b}
V.~Balasubramanian, A.~Lawrence, J.~M.~Magan and M.~Sasieta,
{\em Microscopic Origin of the Entropy of Astrophysical Black Holes,}
\href{https://doi.org/10.1103/PhysRevLett.132.141501}{Phys. Rev. Lett. \textbf{132}, no.14, 141501 (2024)}
\href{https://arxiv.org/abs/2212.08623}{[arXiv:2212.08623 [hep-th]]}.

\bibitem{Climent2024}
A.~Climent, R.~Emparan, J.~M.~Magan, M.~Sasieta and A.~Vilar L{\'o}pez,
{\em Universal construction of black hole microstates,}
\href{https://doi.org/10.1103/PhysRevD.109.086024}{Phys. Rev. D \textbf{109}, no.8, 086024 (2024)}
\href{https://arxiv.org/abs/2401.08775}{[arXiv:2401.08775 [hep-th]]}.

\bibitem{Goel2019}
A.~Goel, H.~T.~Lam, G.~J.~Turiaci and H.~Verlinde,
{\em Expanding the Black Hole Interior: Partially Entangled Thermal States in SYK,}
\href{https://doi.org/10.1007/JHEP02(2019)156}{JHEP \textbf{02}, 156 (2019)}
\href{https://arxiv.org/abs/1807.03916}{[arXiv:1807.03916 [hep-th]]}.

\bibitem{GengJiang2025}
H.~Geng and Y.~Jiang,
{\em Microscopic origin of the entropy of single-sided black holes,}
\href{https://doi.org/10.1007/JHEP04(2025)133}{JHEP \textbf{04}, 133 (2025)}
\href{https://arxiv.org/abs/2409.12219}{[arXiv:2409.12219 [hep-th]]}.

\bibitem{Israel1966}
W.\ Israel,
{\em Singular Hypersurfaces and Thin Shells in General Relativity,}
\href{https://doi.org/10.1007/BF02710419}{Nuovo Cim. B \textbf{44}, 1 (1966)}.

\bibitem{DAlessio2016}
L.~D'Alessio, Y.~Kafri, A.~Polkovnikov and M.~Rigol,
{\em From quantum chaos and eigenstate thermalization to statistical mechanics and thermodynamics,}
\href{https://doi.org/10.1080/00018732.2016.1198134}{Adv. Phys. \textbf{65}, no.3, 239-362 (2016)}
\href{https://arxiv.org/abs/1509.06411}{[arXiv:1509.06411 [cond-mat.stat-mech]]}.

\bibitem{BalasubramanianKang2025}
V.~Balasubramanian, M.~J.~Kang, C.~Cummings, C.~Murdia and S.~F.~Ross,
{\em Purely Greenberger-Horne-Zeilinger-like Entanglement is Forbidden in Holography,}
\href{https://doi.org/10.1103/g5rw-nvnr}{Phys. Rev. Lett. \textbf{136}, no.3, 031602 (2026)}
\href{https://arxiv.org/abs/2509.03621}{[arXiv:2509.03621 [hep-th]]}.

\bibitem{JiangLiu2025}
L.~Jiang and Y.~Liu,
{\em The holographic dual of the GHZ state,}
\href{https://arxiv.org/abs/2508.17898}{[arXiv:2508.17898 [hep-th]]}.

\bibitem{JiangLiu2026}
L.~Jiang and Y.~Liu,
{\em Diving into Booklet Wormholes,}
\href{https://arxiv.org/abs/2603.11459}{[arXiv:2603.11459 [hep-th]]}.

\bibitem{Shen2024}
J.~Y.~Shen, C.~Peng and L.~X.~Li,
{\em Multiway Junction Conditions for Spacetimes with Multiple Boundaries,}
\href{https://doi.org/10.1103/PhysRevLett.133.131601}{Phys. Rev. Lett. \textbf{133}, no.13, 131601 (2024)}.

\bibitem{Shen2024b}
J.~Y.~Shen, C.~Peng and L.~X.~Li,
{\em Multiway junction conditions: Booklets and webs,}
\href{https://doi.org/10.1103/PhysRevD.110.066021}{Phys. Rev. D \textbf{110}, no.6, 066021 (2024)}
\href{https://arxiv.org/abs/2402.00694}{[arXiv:2402.00694 [hep-th]]}.

\Needspace{4\baselineskip}
\bibitem{Shen2026}
J.~Y.~Shen,
{\em Multiway junction conditions: Jackiw-Teitelboim gravity,}
\href{https://arxiv.org/abs/2603.17695}{[arXiv:2603.17695 [hep-th]]}.

\bibitem{LiuWang2025}
Y.~Liu and C.~Y.~Wang,
{\em Energy transport in holographic junctions,}
\href{https://doi.org/10.1007/JHEP10(2025)205}{JHEP \textbf{10}, 205 (2025)}
\href{https://arxiv.org/abs/2506.19553}{[arXiv:2506.19553 [hep-th]]}.

\bibitem{GuoMiao2026}
Y.~Guo and R.~X.~Miao,
{\em Gravity dual of networks,}
\href{https://doi.org/10.1007/JHEP03(2026)112}{JHEP \textbf{03}, 112 (2026)}
\href{https://arxiv.org/abs/2506.21305}{[arXiv:2506.21305 [hep-th]]}.

\bibitem{Banerjee2026}
A.~Banerjee, T.~Kibe, A.~Mukhopadhyay and G.~Policastro,
{\em On decoding the string from interfaces in 2d conformal field theories,}
\href{https://doi.org/10.1007/JHEP07(2026)118}{JHEP \textbf{07}, 118 (2026)}
\href{https://arxiv.org/abs/2511.19592}{[arXiv:2511.19592 [hep-th]]}.

\bibitem{GuoMiao2026b}
Y.~Guo and R.~X.~Miao,
{\em Holographic network, entanglement wedge and traversable parallel universe,}
\href{https://doi.org/10.1007/JHEP07(2026)116}{JHEP \textbf{07}, 116 (2026)}
\href{https://arxiv.org/abs/2601.21206}{[arXiv:2601.21206 [hep-th]]}.

\bibitem{Chakraborty2026}
A.~Chakraborty, T.~Kibe, M.~Molina, A.~Mukhopadhyay and G.~Policastro,
{\em Decoding multiway gravitational junctions in AdS in terms of holographic quantum maps,}
\href{https://arxiv.org/abs/2604.07463}{[arXiv:2604.07463 [hep-th]]}.

\bibitem{deBoer2024}
J.~de Boer, D.~Liska and B.~Post,
{\em Multiboundary wormholes and OPE statistics,}
\href{https://doi.org/10.1007/JHEP10(2024)207}{JHEP \textbf{10}, 207 (2024)}
\href{https://arxiv.org/abs/2405.13111}{[arXiv:2405.13111 [hep-th]]}.

\bibitem{BalasubramanianYildirim2025a}
V.~Balasubramanian and T.~Yildirim,
{\em Observing spacetime,}
\href{https://doi.org/10.1103/p7wt-47rj}{Phys. Rev. D \textbf{113}, no.10, 106034 (2026)}
\href{https://arxiv.org/abs/2509.09763}{[arXiv:2509.09763 [hep-th]]}.

\bibitem{BalasubramanianYildirim2025b}
V.~Balasubramanian and T.~Yildirim,
{\em Nonperturbative toolkit for quantum gravity,}
\href{https://doi.org/10.1103/nlzj-w34h}{Phys. Rev. D \textbf{114}, no.2, 026035 (2026)}
\href{https://arxiv.org/abs/2504.16986}{[arXiv:2504.16986 [hep-th]]}.

\bibitem{Maldacena2003}
J.~M.~Maldacena,
{\em Eternal black holes in anti-de Sitter,}
\href{https://doi.org/10.1088/1126-6708/2003/04/021}{JHEP \textbf{04}, 021 (2003)}
\href{https://arxiv.org/abs/hep-th/0106112}{[arXiv:hep-th/0106112 [hep-th]]}.

\bibitem{HawkingPage1983}
S.~W.~Hawking and D.~N.~Page,
{\em Thermodynamics of Black Holes in anti-De Sitter Space,}
\href{https://doi.org/10.1007/BF01208266}{Commun. Math. Phys. \textbf{87}, 577 (1983)}.

\bibitem{Antonini2023}
S.~Antonini, M.~Sasieta, and B.~Swingle,
{\em Cosmology from Random Entanglement,}
\href{https://doi.org/10.1007/JHEP11(2023)188}{JHEP \textbf{11}, 188 (2023)}
\href{https://arxiv.org/abs/2307.14416}{[arXiv:2307.14416 [hep-th]]}.

\bibitem{Antonini2025}
S.~Antonini, P.~Rath, M.~Sasieta, B.~Swingle and A.~Vilar L{\'o}pez,
{\em The baby universe is fine and the CFT knows it: on holography for closed universes,}
\href{https://doi.org/10.1007/JHEP12(2025)159}{JHEP \textbf{12}, 159 (2025)}
\href{https://arxiv.org/abs/2507.10649}{[arXiv:2507.10649 [hep-th]]}.

\bibitem{Liu2025}
H.~Liu,
{\em ``Filtering'' CFTs at large N: Euclidean Wormholes, Closed Universes, and Black Hole Interiors,}
\href{https://arxiv.org/abs/2512.13807}{[arXiv:2512.13807 [hep-th]]}.

\bibitem{Penington2022}
G.~Penington, S.~H.~Shenker, D.~Stanford and Z.~Yang,
{\em Replica wormholes and the black hole interior,}
\href{https://doi.org/10.1007/JHEP03(2022)205}{JHEP \textbf{03}, 205 (2022)}
\href{https://arxiv.org/abs/1911.11977}{[arXiv:1911.11977 [hep-th]]}.

\end{thebibliography}
\end{document}